%% file: shaw_stappers_weltevrede.tex
\documentclass[fleqn,usenatbib]{mnras}

\usepackage{txfonts}

\usepackage[T1]{fontenc}

\DeclareRobustCommand{\VAN}[3]{#2}
\let\VANthebibliography\thebibliography
\def\thebibliography{\DeclareRobustCommand{\VAN}[3]{##3}\VANthebibliography}

\usepackage{graphicx}	
\usepackage{amsmath}[nointegrals]	
\usepackage{amssymb}	

\usepackage{caption}
\usepackage{subcaption}
\usepackage{pbox}
\usepackage[table]{xcolor}

\title[Long-term pulsar variability]{Long-term rotational and emission variability of 17 radio pulsars}

\author[Shaw, B. et al.]{B. Shaw$^{1}$\thanks{E-mail: benjamin.shaw@manchester.ac.uk}, B. W. Stappers$^{1}$, P. Weltevrede$^{1}$, P. R. Brook$^{2,3}$, A. Karastergiou$^{4,5}$, C. A. Jordan$^{1}$, \newauthor M. J. Keith$^{1}$, M. Kramer$^{6,1}$ and A. G. Lyne$^{1}$
\\
$^{1}$Jodrell Bank Centre for Astrophysics, The University of Manchester, Manchester, M13 9PL, United Kingdom\\
$^{2}$Department of Physics and Astronomy, West Virginia University, P.O. Box 6315, Morgantown, WV 26506, USA\\
$^{3}$Center for Gravitational Waves and Cosmology, West Virginia University, Chestnut Ridge Research Building, Morgantown, WV 26505, USA\\
$^{4}$Astrophysics, University of Oxford, Denys Wilkinson Building, Keble Road, Oxford OX1 3RH, UK\\
$^{5}$Department of Physics and Electronics, Rhodes University, PO Box 94, Grahamstown 6140, South Africa\\
$^{6}$Max-Planck-Institut f\"{u}r Radioastronomie, Auf dem H\"{u}gel 69, D-53121 Bonn, Germany\\
}

\date{Accepted XXX. Received YYY; in original form ZZZ}

\pubyear{2022}

\begin{document}
\label{firstpage}
\pagerange{\pageref{firstpage}--\pageref{lastpage}}
\maketitle

\begin{abstract}
With the ever-increasing sensitivity and timing baselines of modern radio telescopes, a growing number of pulsars are being shown to exhibit transitions in their rotational and radio emission properties. In many of these cases, the two are correlated with pulsars assuming a unique spin-down rate ($\dot{\nu}$) for each of their specific emission states. In this work we revisit 17 radio pulsars previously shown to exhibit spin-down rate variations. Using a Gaussian process regression (GPR) method to model the timing residuals and the evolution of the profile shape, we confirm the transitions already observed and reveal new transitions in 8 years of extended monitoring with greater time resolution and enhanced observing bandwidth. We confirm that 7 of these sources show emission-correlated $\dot{\nu}$ transitions ($\Delta \dot{\nu}$) and we characterise this correlation for one additional pulsar, PSR B1642$-$03. We demonstrate that GPR is able to reveal extremely subtle profile variations given sufficient data quality. We also corroborate the dependence of $\Delta \dot{\nu}$ amplitude on $\dot{\nu}$ and pulsar characteristic age. Linking  $\Delta \dot{\nu}$ to changes in the global magnetospheric charge density $\Delta \rho$, we speculate that $\dot{\nu}$ transitions associated with large $\Delta \rho$ values may be exhibiting detectable profile changes with improved data quality, in cases where they have not previously been observed.
\end{abstract}.




\input{./sections/intro/intro}
\input{./sections/observations/observations}
\input{./sections/method/method}

\input{./sections/results/results}
  \input{./sections/results/0740}

\input{./sections/results/0919}

  \input{./sections/results/1540}
  \input{./sections/results/1642}

  \input{./sections/results/1822}
  \input{./sections/results/1828}

  \input{./sections/results/2035}

  \input{./sections/results/2043} 
  \input{./sections/results/noprofvar}

\input{./sections/discussion/discussion}

    \input{./sections/discussion/discstates}

    \input{./sections/discussion/modtimes}

\input{./sections/discussion/other}

\input{./sections/conclusions/conclusions}

\section*{Acknowledgements}
Pulsar research at the Jodrell Bank Centre for Astrophysics is supported by a consolidated grant from the STFC in the UK. The authors acknowledge the efforts of the staff at the Jodrell Bank Observatory in recording the data used in this manuscript.

\section*{Data Availability}

The data in this paper are available upon reasonable request to the authors.



\bibliographystyle{mnras}
\bibliography{journals,psrrefs,modrefs} 





\bsp	
\label{lastpage}
\end{document}

%% file: sections/intro/intro.tex
\section{Introduction}

Pulsars have long been known for their exceptional rotational stability.  As they lose rotational kinetic energy, their spin-frequencies are observed to gradually reduce over time.  In some cases, a simple timing model consisting of the spin-frequency and its first derivative, as well as the position and dispersion measure, is sufficient to predict pulse times of arrival (TOAs) at the Earth such that the  timing residuals are dominated by measurement error and are spectrally ``white''.  More commonly, timing residuals in ``normal'' (non-recycled) pulsars exhibit significant correlated structures in their residuals (see \cite{hlk10} for a review).  Such structure embodies rotational and/or emission phenomena that are insufficiently described by the timing model.

In some young pulsars (those with characteristic ages less than $\sim$100 kyr), timing irregularites can be dominated by recoveries from glitch events, in which the pulsar undergoes a near discontinuous decrease in its spin period, sometimes followed by a period of recovery. Though not fully understood, these events can be explained by the transfer of angular momentum from an interior superfluid component to the inner crust of the neutron star, forcing it to ``spin-up'' rapidly (see \cite{hm15} for a review of glitch models).  The recovery timescales can extend to months or years after the event. In older pulsars, timing residuals often exhibit a more apparently stochastic wandering of the rotational parameters with respect to the timing model.  This has become known as ``timing noise'' and is widely discussed in the literature (e.g., \citealt{ch80}; \citealt{cor93};  \citealt{pt96}; \citealt{sfal97}; \citealt{qxxw03}; \citealt{cs06}; \citealt{hlk10}; \citealt{lbs+20}) though a conclusive underlying mechanism has not been established.  

\cite{klo+06} showed that the timing noise in the intermittent pulsar PSR B1931+24 can be significantly reduced by including a 50 per cent change in the modelled spin-frequency derivative when the pulsar is not detected. In other words, the pulsar spins down more slowly when it is, or appears to be in a radio silent state.  As the radio emission only accounts for a small fraction of a pulsar's energy budget, this behaviour suggests that such emission variability is part of a global magnetospheric process. \cite{lhk+10} (hereafter LHK) revealed further examples of emission-rotation correlation in 6 pulsars whose pulse shapes (quantified according to a \emph{shape parameter} - a metric of e.g., the pulse width) are associated with particular discrete spin-down rates.  In other words, when a pulsar changes its spin-down rate, changes to the pulse shape (\emph{mode-switching}) also occur.  In a further 11 LHK sources, although strong variations in $\dot{\nu}$ were seen, pulse shape changes were either too subtle to be detected, or do not occur.  

This work serves as an extension to LHK in which we take advantage of a further 8 years of routine timing, yielding up to 47.5 years of rotation history per pulsar. Since LHK, improvements to the pulsar timing program at Jodrell Bank have, in some cases, resulted in a higher attainable signal-to-noise ratio (S/N) due to an increase in the observing bandwidth. Coupled with our improved time resolution, we are sensitive to more subtle changes in the timing behaviour and pulse profile shapes.   We investigate the rotational and emission variability of the 17 pulsars previously studied in LHK and search for correlations between the spin-down rate and the pulse shape.  To quantify their emission and rotational behaviour we use a Gaussian Process Regression (GPR) technique (see \citealt{randw}) that has been utilised for pulsar variability studies by \cite{bkb+14} and \cite{bkj+15} (BKJ hereafter), based on earlier work by \cite{krj+11}.   In BKJ, up to 8 years of data from 168 young pulsars from the Parkes-Fermi timing programme (\citealt{wjm+10}, \citealt{krj+15}) were analysed.  They identified emission-rotation correlation in 9 sources, 8 of which were previously known to exhibit such phenomena. This paper is set out as follows.  In \S2 we describe the observational set-up. \S3 describes how we use the GPR techniques to model variability in our 17 pulsars.  We present our results for each pulsar in \S \ref{gprresults} and in \S \ref{gprdisc} and \S \ref{conc} we discuss the implications of our analysis. 

\begin{table*}
\begin{center}
\begin{tabular}{cccrrrrrrr}
\hline
\hline
    PSR B&PSR J&Epoch (MJD)&$\nu$ (Hz)&$\dot{\nu}$ (10 \textsuperscript{-15} Hz s\textsuperscript{-1})&DM (cm$^{-3}$pc)& RAJ & DECJ &$\tau_{\mathrm{char}}$ (Myr) \\
\hline
    B2148$+$63 &    J2149$+$6329 & 50519 & 2.63 &  $-$1.18 &  129.2 & 21:49:58.62 & $+$63:29:43.80 & 35.8  \\
       -      &    J2043$+$2740 & 54866 & 10.40 & $-$134.57 & 21.0 & 20:43:43.50 & $+$27:40:56.40 &  1.2 \\ 
    B2035$+$36 &    J2037$+$3621 & 50085 & 1.62 & $-$11.82 &   93.6 & 20:37:27.44 & $+$36:21:24.10 &  2.2 \\
    B1929$+$20 &    J1932$+$2020 & 54493 & 3.73 & $-$50.87 &   211.2 &  19:32:08.02 & $+$20:20:46.41 & 1.0 \\
    B1907$+$00 &    J1909$+$0007 & 51194 & 0.98 & $-$5.33 &    112.7 & 19:09:35.26 & $+$00:07:57.67 & 2.9 \\
    B1903$+$07 &    J1905$+$0709 & 53751 & 1.54 & $-$1.18 &    245.3 & 19:05:53.62 & $+$07:09:19.40 & 2.1 \\
    B1839$+$09 &    J1841$+$0912 & 54125 & 2.62 & $-$7.50 &    49.2 & 18:41:55.96 & $+$09:12:07.35 &  5.5 \\
    B1828$-$11 &    J1830$-$1059 & 53311 & 2.47 &  $-$365.49 & 161.5 & 18:30:47.58 & $-$10:59:29.33 & 0.1 \\
    B1826$-$17 &    J1829$-$1751 & 54146 & 3.26 &  $-$50.88 &  217.0 & 18:29:43.14 & $-$17:51:04.05 & 0.8 \\
    B1822$-$09 &    J1825$-$0935 & 52740 & 1.30 &  $-$88.57 &  19.4 & 18:25:30.61 & $-$09:35:21.37 &  0.2 \\
    B1818$-$04 &    J1820$-$0427 & 47963 & 1.67 &  $-$17.70 & 84.4 & 18:20:52.60 & $-$04:27:38.34 &  1.5  \\
    B1714$-$34 &    J1717$-$3425 & 54069 & 1.52 &  $-$22.27 &  587.7 & 17:17:20.26 & $-$34:25:00.05 & 1.1  \\
    B1642$-$03 &    J1645$-$0317 & 54088 & 2.58 &  $-$1.18  &  35.8 & 16:45:02.04 & $-$03:17:58.26 &  3.5  \\
    B1540$-$06 &    J1543$-$0620 & 53873 & 1.41 &  $-$1.75  &  18.4 & 15:43:30.16 &  $-$06:20:45.25 &  12.8 \\
    B0950$+$08 &    J0953$+$0755 & 51541 & 3.95 &  $-$3.59  &  3.0  &  09:53:09.30 & $+$07:55:35.94 &  17.5 \\
    B0919$+$06 &    J0922$+$0638 & 49653 & 2.32 &  $-$73.93 &  27.3 & 09:22:14.02 & $+$06:38:23.12 &  0.5  \\
    B0740$-$28 &    J0742$-$2822 & 52014 & 6.00  & $-$604.40 & 73.8 & 07:42:49.06 & $-$28:22:43.72 &  0.2  \\
\hline
\end{tabular}
\end{center}
\caption[]{\label{gprpulsars}Properties of the pulsars used in this work. We list the spin frequency $({\nu})$ and $(\dot{\nu})$, and pulsar's position as measured at the epoch shown in the third column.  We also show the dispersion measure (DM), and the spin-down (characteristic) age $\tau_{\mathrm{char}}$.}
\end{table*}


%% file: sections/observations/observations.tex
\section{Observations} \label{gprobs}

Observations were mostly carried out using the 76-m Lovell Telescope at the Jodrell Bank Observatory (JBO) with supplementary observations made using the 38x25 m Mark II telescope, also at JBO.  Pre-2009 observations were made with an analogue filterbank (AFB) using a 32 MHz bandwidth centred on 1400 MHz (L-Band), except for the period between November 1997 and May 1999 when a 96 MHz bandwidth was used.  Supplementary observations were made at centre frequencies of 400, 610 and 925 MHz over a 4-8 MHz bandwidth. Depending on the pulsar the profile was generated by online folding onto either 400 or 512 bins per pulse period. From 2009 onwards a digital filterbank (DFB) was used wherein the majority of observations were centred on 1520 MHz over a 384 MHz bandwidth with a small number of observations centred on 1400 MHz using a 512 MHz or a 128 MHz bandwidth. In this case the pulses were folded onto 1024 bins per pulse period.  Pulsars are typically observed for durations of 300-800 seconds. DFB data are folded into sub-integrations of 10s duration and 768 frequency channels. 256 channels were used pre-2009.  Detailed descriptions of the backends used here can be found in \cite{sl96} and \cite{hlk+04} (AFB), and \cite{mhb+13} (DFB). 

To excise radio frequency inteference (RFI) we first apply a median-filtering algorithm, followed by manual removal of any remaining RFI from affected frequency channels or sub-integrations. 

A single integrated pulse profile is created for each observation epoch by summing the data in all frequency channels, sub-integrations and polarisations. In order to compute a topocentric time of arrival (TOA), using \textsc{psrchive} \citep{hsm04} we cross-correlate the integrated profile with a high S/N template profile (generated from von Mises functions fitted to high S/N observations) which represents the expected shape of the observed profile for a particular observing frequency and bandwidth.  

The measured TOAs are converted to barycentric arrival times. Timing residuals are then formed by subtracting the measured arrival times from those predicted by a timing model of the pulsar which contains its rotational and astrometric parameters. For this, we make use of the pulsar timing suite \textsc{tempo2} \citep{hlk10}.

%% file: sections/method/method.tex
\section{Methodology} \label{methodology}

\subsection{Data preparation}

\begin{figure}
    \includegraphics[width=1\columnwidth]{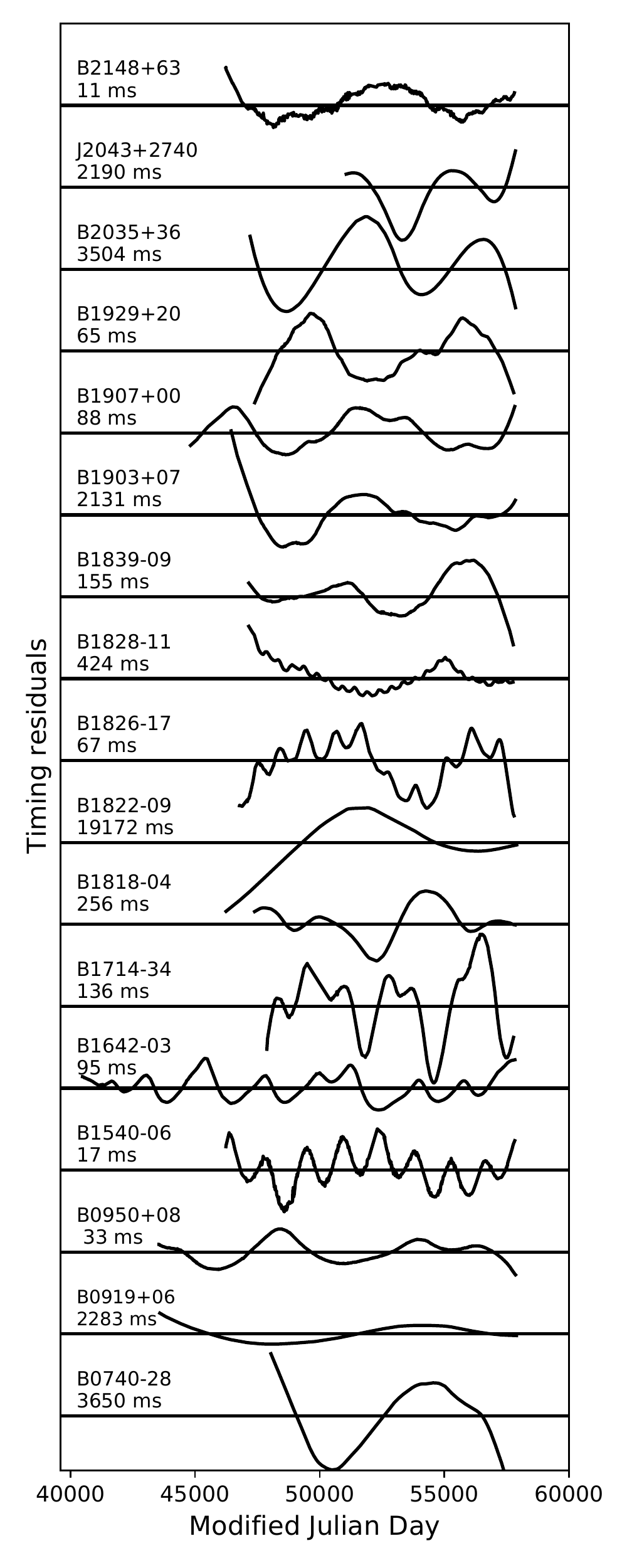} 
    \caption[Timing residuals relative to simple slow-down models for pulsars that exhibit timing noise]{Up to 49 years of timing residuals following the subtraction of a spin-down model comprising the spin-frequency and its constant first time derivative $\dot{\nu}$ (see Table \ref{gprpulsars}). The approximate peak-to-peak values of the residuals are shown beneath the pulsar names.}
    \label{resids}
\end{figure}

To compare pulse shapes we need to ensure that we are comparing like-with-like. As pulse shapes can have a strong frequency dependence (e.g., \citealt{hr86}, \citealt{phs+16}), the data used for profile variability analysis must generally share the same centre frequency and bandwidth. For this reason, we undertake our profile variability studies seperately for each backend used. For the AFB data (until 2009), we use profiles observed at 1400 MHz with a 32 MHz-wide band. Data obtained using the 96 MHz-wide band were only used where it was confirmed by manual inspection that the increased bandwidth did not introduce sharp changes in the observed pulse profile shape. Similarly, in the DFB era, we use only the profile observed at a centre frequency of 1532 MHz over a 384 MHz wide band. 

For pulse TOAs, it is not the case that data need to be excluded based on the use of different backends and receivers.  For example, if a pulsar is observed at two different frequencies (at which there is no difference in the pulse profile shape), this will give rise to a constant offset between the two sets of TOAs. For this reason fewer observations are used for monitoring profile variability than for spin-down variability.  


Figure \ref{resids} shows the timing residuals for pulsars used in this study after fitting for the pulsar's spin frequency, spin-down rate, and position. We list the rotational and other properties of the pulsars in our sample in Table \ref{gprpulsars}. In addition, both PSRs B0740$-$28 and B1822$-$09 have undergone large glitches and so we also include the glitch parameters in the timing models for those pulsars. 

\subsection{Gaussian process models of pulsar variability}

In order to model the variations in pulse shape and $\dot{\nu}$ in the 17 pulsars, we closely follow the procedures outlined comprehensively in BKJ. We briefly summarise the process below. 

\subsubsection{Pulse shape variations}

\begin{figure}
    \includegraphics[width=1\columnwidth]{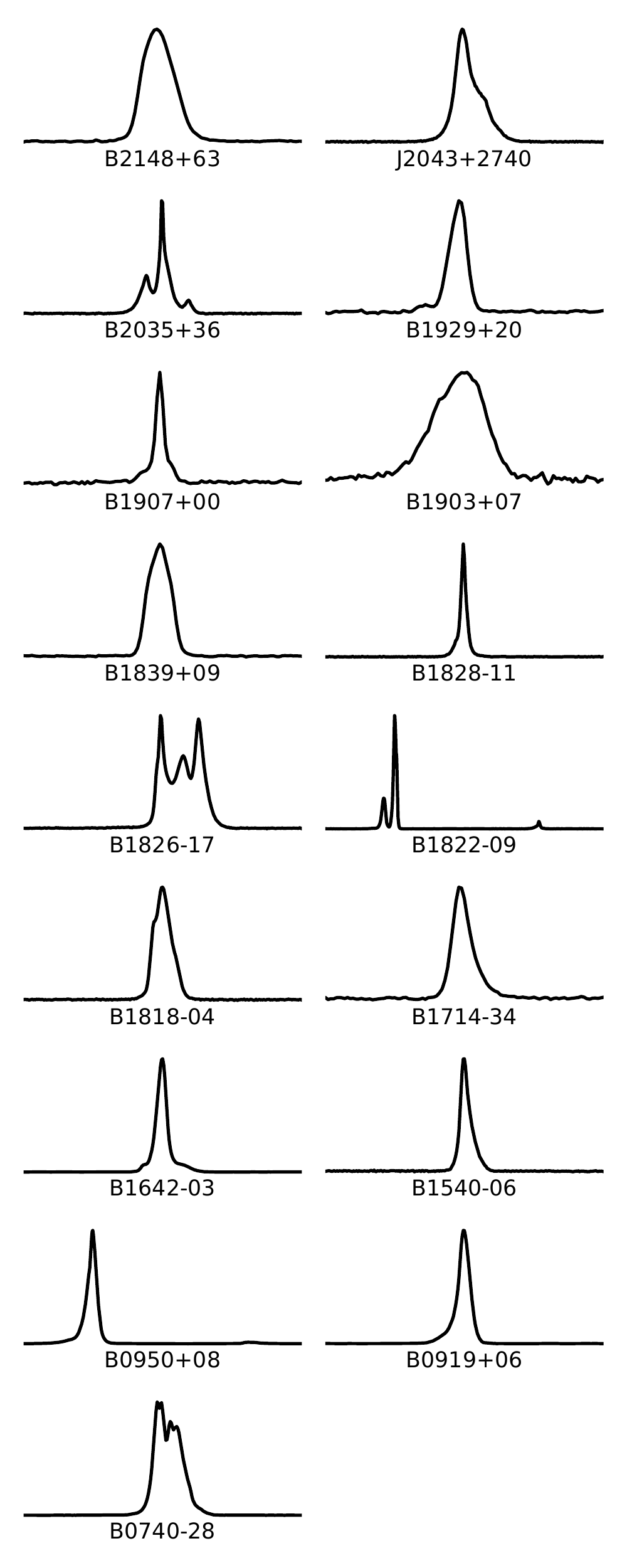} 
    \caption{The median $\sim$1.5 GHz pulse profiles of the 17 pulsars studied in this work. The profiles are formed from all individial integrated profiles for each source, that were observed after 2009 (see \S\ref{gprobs}). For each profile, 90 degree of rotation (a quarter of one period) is shown. This is with the exception of PSRs B0950$+$08 and B1822$-$09, of which, due to the presence of interpulses, one full rotation is shown.}
    \label{profiles2}
\end{figure}

For each pulsar we initially remove, by inspection, any observations which have a noticeably low signal-to-noise ratio (S/N) or are clearly distorted due to either instrumental failures or significant levels of RFI.  The off-pulse mean of each profile is then subtracted from all phase bins to ensure the noise-level between successive profiles has a mean value of zero. The individual profiles are then normalised by the mean on-pulse flux density.  This leaves us sensitive to changes in pulse shape, but insensitive to changes to flux density as our data are not flux-calibrated. Pulse profiles from each observation are cross-correlated and rotated to a common phase such that they are ``phase-aligned''. These profiles are added together to form a high S/N ratio template profile. The template is then subtracted from the individual profiles leaving behind a \emph{profile residual} which, if the integrated profile remains stable across all epochs, should approximate white noise.  An excess or lack of power at any pulse phase therefore, represents a deviation from the median profile shape at that epoch. The template profiles for the 17 pulsars studied in this work are shown in Figure \ref{profiles2}.

Within a manually specified \emph{on-pulse} region, we compute a Gaussian Process (GP) to model the behaviour of the profile residuals at each phase bin over all epochs. As the shape of a profile can change dramatically from observation to observation, we employ a variant of the Mat\'{e}rn covariance function \citep{randw} to model the covariance between the power values in each phase bin, as it is capable of modelling sharp features in the profile residuals. The Mat\'{e}rn covariance is a function of two \emph{hyperparameters}, $\lambda$ and $\sigma^2$ which respectively describe the smoothness and the variance of the function that describes the profile variations (see \cite{randw}; \cite{brook15}; \cite{shaw18} for detailed descriptions of the hyperparameters). We also use an additional white noise covariance function (with hyperparameter $\sigma_\mathrm{N}^{2}$) to model the uncertainty on the profile residuals.  For each phase bin, the values of $\lambda$, $\sigma^2$ and $\sigma_\mathrm{N}^{2}$ are optimised using a maximum likelihood function, resulting in an analytical function describing the profile residual. This allows us to plot a \emph{variability map} (Section \ref{gprresults}) to show the behaviour of the profile with respect to the template as a function of phase and time. 

\subsubsection{$\dot{\nu}$ variations}

In order to quantify a time-variable $\dot{\nu}$, we use GPR to fit a continous function to the timing residuals, the second derivative of which, yields the spin-down evolution (see \citealt{ksj13}).  To calculate $\dot{\nu}(t)$ we employ the squared exponential (SE) covariance function, as it has infinitely many derivatives. Though the SE covariance function has a different form to the Mat\'{e}rn covariance, it shares the same hyperparameters. As with the pulse shape variations technique, we use this covariance function in combinations with a white noise covariance function in order to model noise in the timing residuals.

The value of $\dot{\nu}$ is only inferred on days corresponding to observation epochs. For several of our pulsars a single covariance function was insufficient to fully model the timing residuals. This implies that the variations in $\dot{\nu}$ are not occurring over a single characteristic timescale.  In such circumstances, a second additive covariance function resulted in a better fit to the data. In these cases, there are five hyperparameters to optimise. The two lengthscales $\lambda$, the two signal variances $\sigma^2$ and a single noise variance $\sigma^2_\mathrm{N}$.

In order to qualitatively examine the correlation between the profile shape and $\dot{\nu}$, we compute the Spearman-Rank correlation coefficient (SRCC) between the profile residuals in each on-pulse phase bin and the $\dot{\nu}$ time series. This requires that the data in the two time series are equally sampled, therefore we recompute the values of $\dot{\nu}$ at daily intervals to correspond with the inference interval of the profile residual, in line with BKJ. A lag is applied in either direction between the two time series in order to measure how the correlation changes and to identify any periodicities. Where a correlation is identified between $\dot{\nu}$ and the profile shape, we also show, where possible, examples of the profile in each of the states.  Where the differences between profiles are not discernible by inspecting differences between individual observations, we average together all profiles observed when the value of $\dot{\nu}$ was above and below a threshold value for that pulsar. We also apply this to cases where no profile variations were detectable from the GPR variability maps in order to examine whether there are subtle average differences in the profile in each $\dot{\nu}$ state.

%% file: sections/results/results.tex
\section{Results} \label{gprresults}

\begin{table*}
\begin{center}
\begin{tabular}{lccccccccc}
\hline
\hline
PSR & \pbox{10cm}{Dataspan (MJDs)} & $N_{\mathrm{TOA}}$ & \pbox{10cm}{$\langle{\Delta t}\rangle$  (days)} & \pbox{10cm}{$\sigma_1^2$ (s)} & \pbox{10cm}{$\lambda_1$} &  $\sigma_2^2$ (s) & \pbox{10cm}{$\lambda_2$} & $\sigma_N^2$ (s) \\
\hline
     B2148$+$63 & 46237 $-$ 57632 & 594 & 19 & $2.9 \times 10^{-5}$ & 448  & - & - & $4.3 \times 10^{-8}$   \\ 
     J2043$+$2740 & 51061 $-$ 57848 & 567 & 12 & $1.9 \times 10^{-1}$ & 324 & $3.1 \times 10^{-7}$ & 39  & $2.5 \times 10^{-12}$ \\
     B2035$+$36 & 47389 $-$ 57785 & 524 & 30 & $1.4 \times 10^{-5}$ & 82  & $2.3 \times 10^{-0}$ & 778 & $1.1 \times 10^{-7}$   \\ 
     B1929$+$20 & 47392 $-$ 57779 & 394 & 26 & $3.0 \times 10^{-3}$ & 333  & - & - & $2.8 \times 10^{-8}$  \\
     B1907$+$00 & 44818 $-$ 57821 & 410 & 32 & $5.4 \times 10^{-4}$ & 430  & - & - & $1.3 \times 10^{-7}$  \\ 
     B1903$+$07 & 46301 $-$ 57844 & 407 & 28 & $1.8 \times 10^{-1}$ & 342  & - & - & $7.4 \times 10^{-6}$  \\ 
     B1839$+$09 & 47616 $-$ 54125 & 437 & 23 & $1.1 \times 10^{-3}$ & 235 & - & - & $7.9 \times 10^{-8}$  \\ 
     B1828$-$11 & 49142 $-$ 57818 & 1334 & 7 & $2.1 \times 10^{-1}$ & 109 & - & - & $7.7 \times 10^{-8}$  \\ 
     B1826$-$17 & 46781 $-$ 57792 & 561 & 20 & $1.3 \times 10^{-3}$ & 435 & $2.3 \times 10^{-6}$ & 96 & $1.6 \times 10^{-8}$ \\
     B1822$-$09 & 46242 $-$ 57898 & 1289 & 9 & $8.2 \times 10^{1}$ & 221 & - & - & $5.2 \times 10^{-7}$ \\
     B1818$-$04 & 47388 $-$ 57837 & 508 & 21 & $1.1 \times 10^{-6}$ & 136 & $3.4 \times 10^{-2}$ & 757 & $2.5 \times 10^{-8}$ \\
     B1714$-$34 & 47780 $-$ 57771 & 266 & 37 & $1.5 \times 10^{-2}$ & 711 & $6.2 \times 10^{-6}$ & 153 & $2.3 \times 10^{-7}$   \\ 
     B1642$-$03 & 40485 $-$ 57826 & 1179 & 15 & $2.8 \times 10^{-4}$ & 257 & - & - & $3.2 \times 10^{-8}$   \\
     B1540$-$06 & 46434 $-$ 57821 & 717 & 16 & $4.0 \times 10^{-4}$ & 497 & - & - & $5.7 \times 10^{-8}$  \\ 
     B0950$+$08 & 43549 $-$ 57850 & 1220 & 12 & $4.2 \times 10^{-5}$ & 526 & - & - & $2.7 \times 10^{-9}$   \\ 
     B0919$+$06 & 43586 $-$ 57890 & 1156 & 12 & $1.1 \times 10^{-1}$ & 170  & - & - & $3.6 \times 10^{-8}$ \\
B0740$-$28 & 48042 $-$ 58181 & 1848 & 5 & $3.0 \times 10^{-6}$ & 46  & $8.8 \times 10^{-1}$ & 660 & $4.4 \times 10^{-9}$ \\

\hline
\end{tabular}
\end{center}
\caption[]{Model parameters for each pulsar.  The optimised variance $\sigma_i^2$, kernel lengthscale, $\lambda_i$ and noise variance $\sigma_N^2$  are shown for each source as well as the number of TOAs and average cadence over the dataspan $\langle{\Delta t}\rangle$. Where a single kernel was sufficient to model the timing residuals, the fields $\sigma_2^2$ and $\lambda_2$ are left empty. }
\label{cov}
\end{table*}

\begin{figure}
    \includegraphics[width=1\columnwidth]{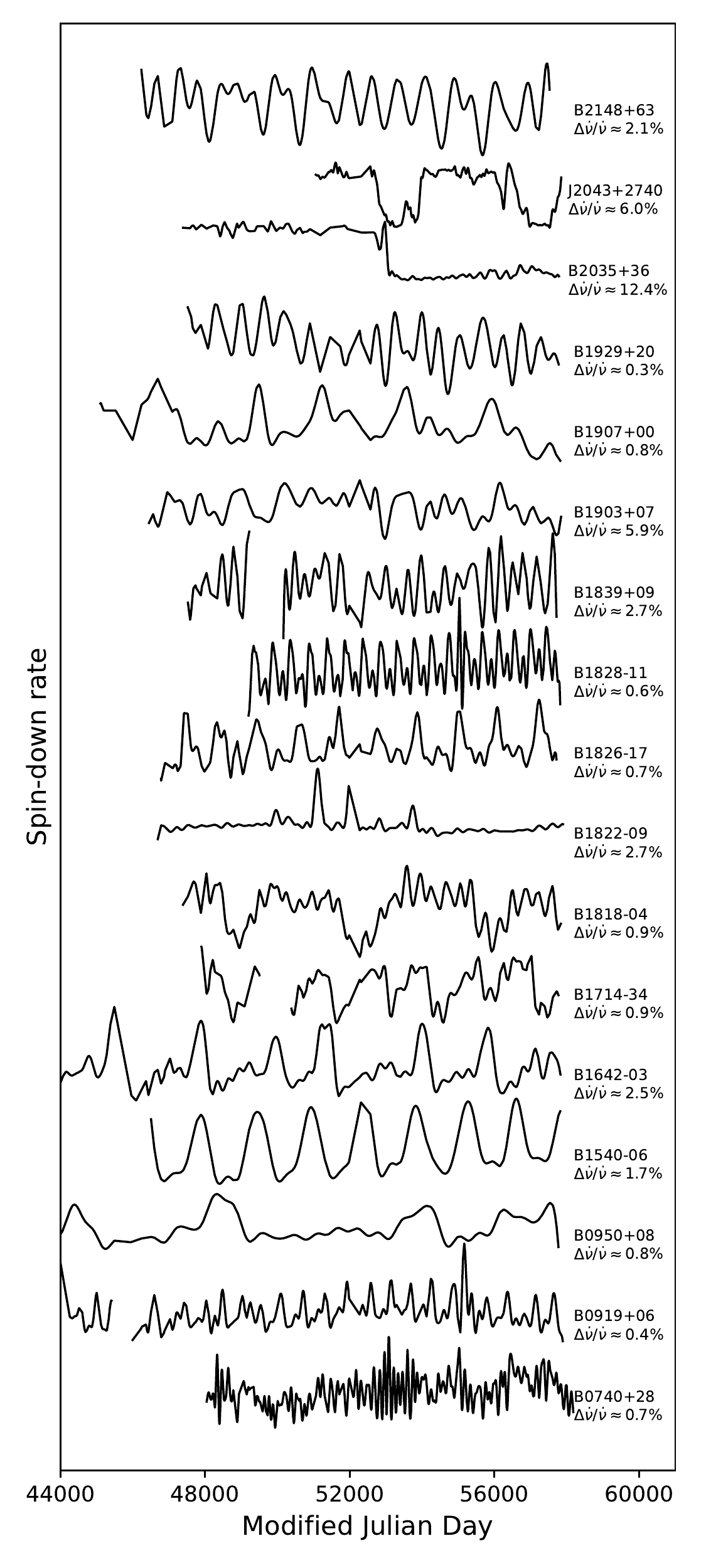}
    \caption{Variations in $\dot{\nu}$ of the pulsars studies in this work. A downward deflection represents an increase in the magnitude of $\dot{\nu}$. The approximate peak-to-peak fractional amplitude of the variations ($\Delta \dot{\nu} /  \dot{\nu}$) are shown. Gaps denote long period where no observations took place (see text for details).}
    \label{nudots_prof1}
\end{figure}

We have applied the techniques outlined above and in BKJ to the 17 $\dot{\nu}$-variable pulsars in LHK.  Table \ref{cov} lists the inputs and model parameters for each pulsar used in the search for $\dot{\nu}$ variations.  The majority of pulsars required only a single covariance function to model their timing residuals which are shown in Figure \ref{resids}.  Where pulse shape changes are identified, we present variability maps showing the behaviour of the profile over time and phase, relative to the template profiles shown in Figure \ref{profiles2}. Maps for the AFB and DFB datasets are shown separately. The $\dot{\nu}$ time series is shown beneath each source's variability map allowing the identification of any emission-rotation correlation (see Figure \ref{0740_maps} caption for more information).  The $\dot{\nu}$ variations of all 17 pulsars are shown in Figure \ref{nudots_prof1}. We show the SRCC maps for all pulsars in which correlated emission and spin-down variability was revealed, in Figure \ref{srccgroup}, along with examples of the pulse profiles in each emission state. In order to identify periodicities in the $\dot{\nu}$ variations in the presence of unevenly spaced observations and compare the values to those noted in LHK, we use the Lomb-Scargle spectral analysis tools provided by the astropy.stats package (see \citealt{rtg+13}). The resulting power spectra are shown in Figure \ref{LS}.

%% file: sections/results/0740.tex
\subsection{Pulsars with emission-rotation correlation}
\subsubsection{PSR B0740$-$28} \label{0740sec}

PSR B0740$-$28 is known to show a complicated relationship between the shape of the pulse profile and the value of $\dot{\nu}$. The profile exhibits two distinct shape extremes (see Figure \ref{srccgroup}(A), panel ii) though many observations show a clear mixing between the two states. The profile has a similar width in each of the two states, though in the more common of the two, the trailing component has a notably lower flux density relative to the leading component. In the less common state, the profile is much more symmetric. Of the LHK sample of pulsars, PSR B0740$-$28 is the source whose $\dot{\nu}$ value showed the most rapid modulations. It also shows the most erratic relationship between $\dot{\nu}$ and the profile shape.  The LHK analysis showed that the rate of mode-switching is highly variable over time and not always periodic.  \cite{ksj13} showed that the profile shape became particularly well correlated with $\dot{\nu}$ following the MJD 55022 glitch.

\begin{figure*}
  \includegraphics[width=2\columnwidth]{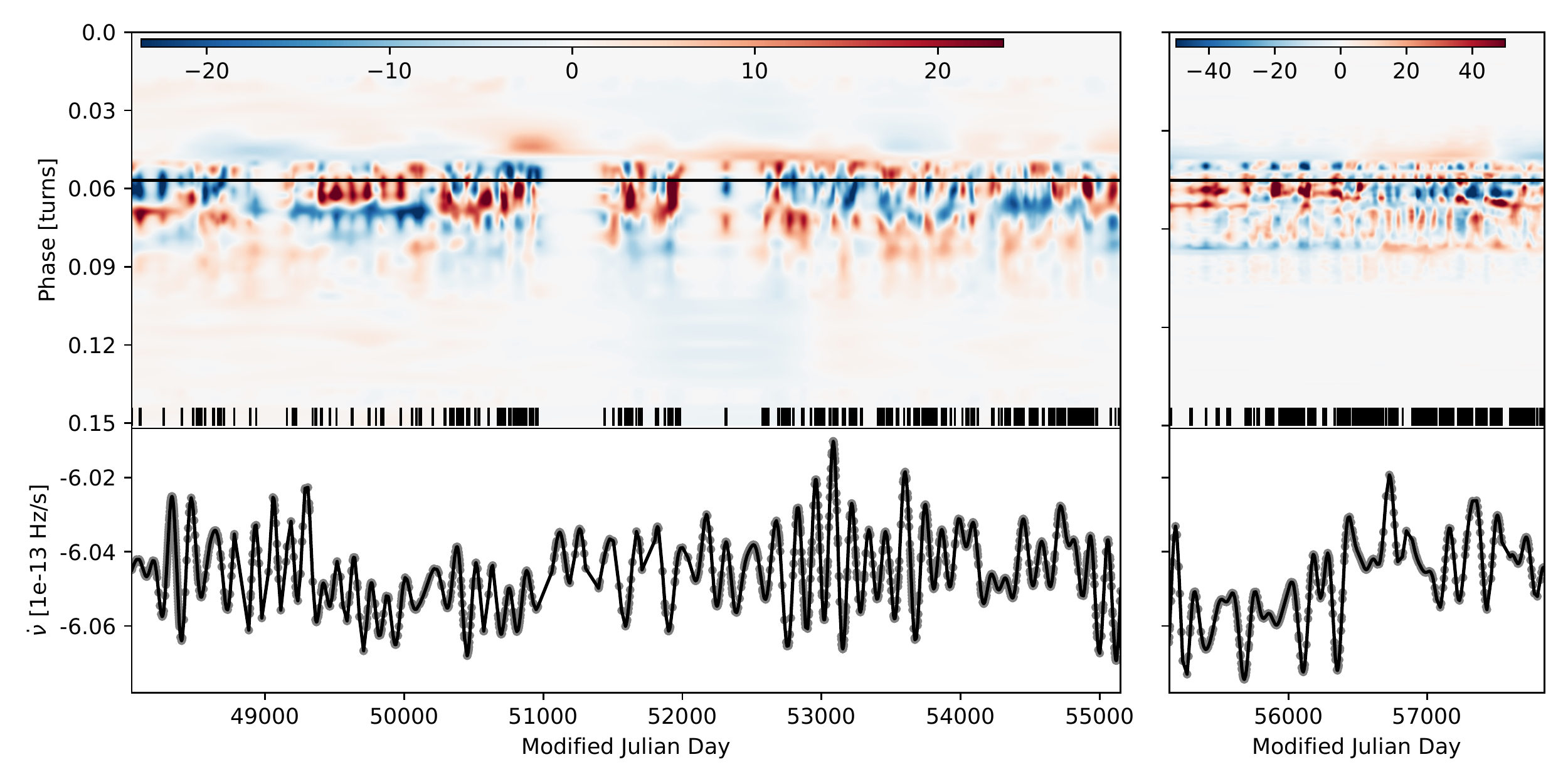}\hspace*{-10mm}
  \caption[Profile and $\dot{\nu}$ variations in PSR B0740$-$28]{\label{0740_maps}Profile variation maps and $\dot{\nu}$ variations in PSR B0740$-$28 for the pre-2009 analogue filterbank (AFB) data (left plot) and the post-2009 digital filterbank (DFB) data (right panel).  The variation maps show the pulse profile variability. Red regions represent an excess in power at that pulse phase relative to the median template profile and blue represents a deficit. The short black lines denote observation epochs and the solid horizontal line is the location of the pulse profile peak in the template profile.  The units (shown in the colourbar) are the median of the standard deviation of all off-pulse regions across the dataset.}
\end{figure*}

The $\dot{\nu}$ variations over time are shown in the lower panels of Figure \ref{0740_maps}. Overall, the periodicity of the variations is rapid, with a maximum occurring roughly every 50-100 days. The value of $\dot{\nu}$ oscillates about a mean value of $-6.05 \times 10^{-13}$ Hz s\textsuperscript{$-1$} with a mean peak-to-peak fractional amplitude of $\Delta \dot{\nu} / \dot{\nu} \approx 0.7$ per cent. However, $\Delta \dot{\nu} / \dot{\nu}$ is relatively low between MJDs 50000 and 52500, then becoming somewhat higher again until MJD 54000. Within these latter dates, the periocity of the $\dot{\nu}$ variations is also much more well defined. Following MJD 54000, $\Delta \dot{\nu} / \dot{\nu}$ apparently reduces again.

The upper panels of Figure \ref{0740_maps} show the profile variability of PSR B0740$-$28 for the AFB and DFB datasets.  As noted by LHK, the level of correlation between the profile shape and $\dot{\nu}$ is variable however it is most clear in the higher time-resolution DFB data in which the trailing part of the profile is brightest when $\dot{\nu}$ assumes its largest values. This is particularly notable in the $|\dot{\nu}|$ maxima (dips in the $\dot{\nu}$ time series) near MJDs 55900, 56125 and 56375.  The correlation is less clearly seen in the AFB data, though we note that between MJDs 49000 and 51000, the central and trailing parts of the profile appear \emph{anti-correlated} with the value of $\dot{\nu}$. This exchange between correlation and anti-correlation was also noted in BKJ. 

The lower panel of Figure \ref{srccgroup}(A) shows the correlation between $\dot{\nu}$ and the pulse shape in the DFB era as a function of pulse phase. The map illustrates the complex relation between the profile shape and $\dot{\nu}$. The central blue region (at $\sim$ phase 0.035), representing the central region of the pulse profile) denotes a region of strong anti-correlation between the two time series maximising at zero lag indicating that $\dot{\nu}$ has a larger magnitude when that region of the profile is brightest.  Similary a region of strong correlation near phase 0.057 shows that the trailing edge region of the profile is brighter when the magnitude of $\dot{\nu}$ is small.

%% file: sections/results/0919.tex
\subsubsection{PSR B0919$+$06}

PSR B0919$+$06 is a bright pulsar with spin-period of 430 ms. The profile (see Figure \ref{profiles2}) normally consists of a single sharp pulse whose leading edge increase is slightly shallower than its trailing edge descent.  Sporadically punctuating the typical emission is a "flare" state identified by \cite{zrw06} in which emission appears earlier in phase for 5-15 seconds every several thousand pulses. Though clear variations in $\dot{\nu}$, comprising consecutive major and minor peaks, have been described in other works (LHK; \citealt{shab10}). \cite{psw14}  demonstrated a clear correlation between $\dot{\nu}$ and the pulse shape using 30 years of Jodrell Bank data.  They also showed that the flare state was not correlated with $\dot{\nu}$, implying that it is not actuated by global effects in the magnetosphere.  \cite{shab10} reported the detection of a very large glitch occurring on MJD 55140 in which the spin-frequency of B0919$+$06 underwent a fractional change  $\Delta \nu / \nu = 1.3 \times 10^{-6}$, though no coincident profile variations have been reported.

The value of $\dot{\nu}$ over time is shown in Figure \ref{nudots_prof1} where clear, repeating double-peaked modulations are seen, bearing resemblence to PSR B1828$-$11 (Section \ref{1828} and Figure \ref{nudots_prof1}). The double-peaked nature was less clear prior to MJD $\sim$50000 where $\dot{\nu}$ was more erratic but with some clear major and minor peaks.  The value of $\dot{\nu}$ modulates about the mean with a peak-to-peak fractional amplitude of $\Delta \dot{\nu} / \dot{\nu} \approx 0.4$ per cent. Using the entirety of our $\dot{\nu}$ data for PSR B0919$+$06 we find that the Lomb-Scargle spectrum peaks at 0.630(2) yr\textsuperscript{-1} corresponding to an average cycle time of $\sim$580 days (see Figure \ref{LS}).

\begin{figure}
    \includegraphics[width=1.0\columnwidth]{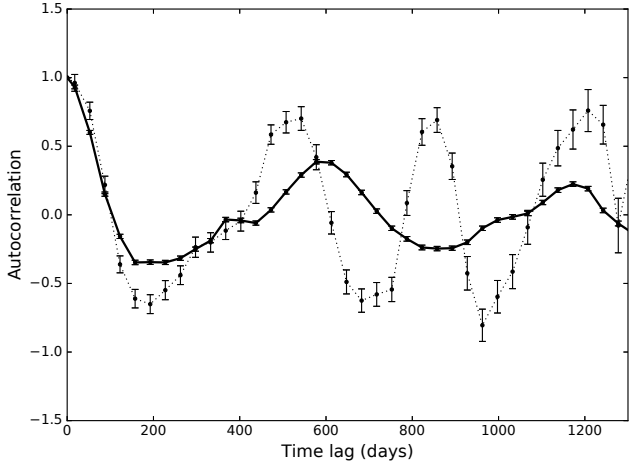}
    \caption[Autocorrelation functions (ACF) of the $\dot{\nu}$ variations in PSR B0919$+$06]{\label{acf0919}Autocorrelation functions (ACF) of the $\dot{\nu}$ variations in PSR B0919$+$06 up to MJD 52000 (solid line) and from MJD 52000 to MJD 56500 (dotted line).  The locations of the first maxima in the ACF denote the values of the strongest periodicity in the datasets.  The earlier data shows a periodicity of $\sim$630 days which drops to $\sim$570 days in the later data.}
\end{figure}

\cite{psw14} noted a change in the modulation timescale of $\dot{\nu}$ before and after a telescope outage near MJD 52000.  By computing separate autocorrelation functions of data prior to and following MJD 52000 they showed that the periodicity of the $\dot{\nu}$ modulations (the distance between consecutive major or minor peaks) was 80 days shorter in the latter period.  By computing autocorrelation functions for the same two time windows using our Gaussian process inferred values of $\dot{\nu}$ we find periodicities of $\sim$630 days prior to MJD 52000 and $\sim$570 days for 52000 $<$ MJD $<$ 56500 (denoted by the first maxima in the autocorrelation functions for each data segment, Figure \ref{acf0919}). These values are very similar to those quoted in \cite{psw14} for the same time periods confirming that the $\dot{\nu}$ modulation timescale became shorter after MJD 52000. To compute the errorbars on the autocorrelation functions, we followed the same procedure outlined in \cite{psw14}.

\begin{figure}
    \includegraphics[width=1.0\columnwidth]{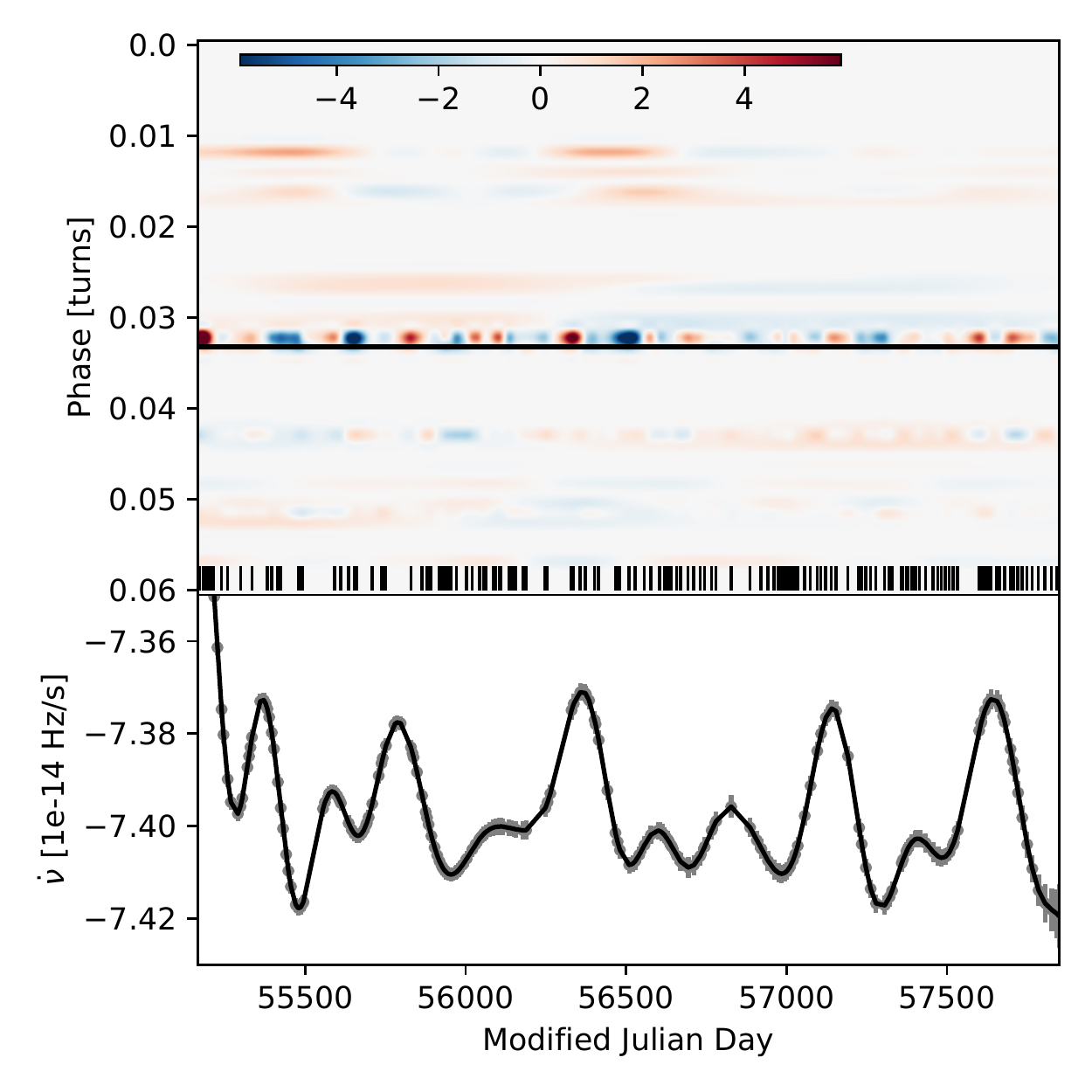}
    \caption[Profile and $\dot{\nu}$ variations in PSR B0919$+$06]{\label{maps_dfb_0919}Profile variation maps and $\dot{\nu}$ variations in PSR B0919$+$06. Only the DFB data is shown here.  No profile variations were seen in the lower quality AFB data.  As Figure \ref{0740_maps} otherwise.}
\end{figure}

In the higher quality DFB data (Figure \ref{maps_dfb_0919}), some clear correlation exists between the profile shape and the spin-down rate. \cite{psw14} noted that pulse profile variations were resolvable in the higher quality DFB data recorded after MJD 55140 and that these variations correlate with $\dot{\nu}$.  Peaks/troughs in $\dot{\nu}$ (corresponding to weak/strong spin-down respectively) appear to correlate with an excess/deficit of emission just ahead of the pulse peak (near phase 0.032) - especially prior to MJD 56500. This is particularly apparent at the $|\dot{\nu}|$ maxima at MJDs 55000 and 56300 and the $|\dot{\nu}|$ minima at MJD 56200 and 56500.  The strong peak at MJD 55750 appears to advance a brightening of emission at phase 0.032 by $\sim$70 days whereas the $\dot{\nu}$ spike at MJD 55250 lags the brightening. These offsets are likely due to a relatively low cadence at these times. Figure \ref{srccgroup}(B) shows the correlation between the pulse shape and $\dot{\nu}$ as a function of pulse phase for the DFB data. We note the strongest correlation occurs at lag zero corresponding to part of the profile near the pulse peak in Figure \ref{maps_dfb_0919} (near phase 0.032). The correlation maximises again at a lag of $\pm 500$ days roughly corresponding to the time between the first three strong peaks in $\dot{\nu}$ in the DFB data.  

%% file: sections/results/1540.tex
\subsubsection{PSR B1540-06}

PSR B1540$-$06 is a bright pulsar with a single-component asymmetric profile characterised by a sharp rise at the leading edge followed by a more gradual decay at the trailing edge and occupies $\sim$2 per cent of a pulse period (Figure \ref{profiles2}). LHK noted very low amplitude variability in the trailing edge of the pulse which cycles with variations in $\dot{\nu}$.  The variations in $\dot{\nu}$ are shown in Figure \ref{nudots_prof1} and exhibit remarkable regularity. The value of $\dot{\nu}$ cycles about the mean value with $\Delta \dot{\nu} / \dot{\nu} \approx 1.7$ per cent.  The variations are characterised by strong peaks occurring roughly every 1500 days.  Low amplitude minor peaks are seen between most of the major peaks. Analysis of the Lomb-Scargle spectral power (Figure \ref{LS}) shows a strong spike at 0.250(1) yr\textsuperscript{-1}, corresponding to a cyle time of 1455 days.

\begin{figure*}
    \includegraphics[width=2.0\columnwidth]{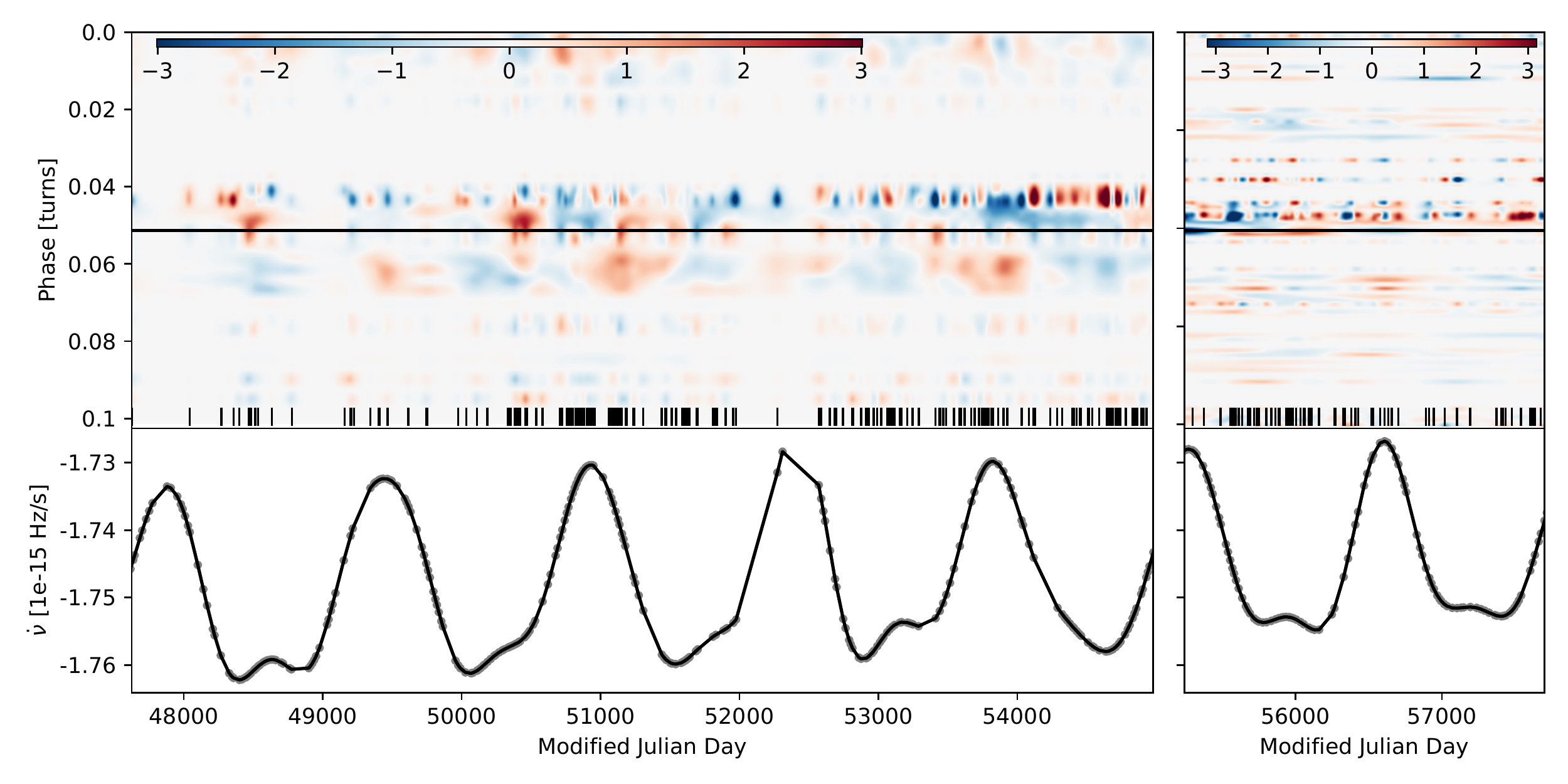} 
    \caption[Profile and $\dot{\nu}$ variations in PSR B1540$-$06]{Profile variation maps and $\dot{\nu}$ variations in PSR B1540$-$06.  As Figure \ref{0740_maps} otherwise.}
    \label{1540_maps}
\end{figure*}

The profile variability maps are shown in Figure \ref{1540_maps}. Clear excesses in power occur at the trailing edge (near phase 0.06) in the AFB map when the pulsar's spin-down rate is weakest (MJDs 49500, 51000, 53800).  The weak spin-down value near MJD 52300 appears less well correlated with the power increase in the trailing edge due to the pulsar being observed with a lower cadence near that time.  We note that in the DFB data, the correlation between $\dot{\nu}$ and the profile shape is less clear as only two peaks in $\dot{\nu}$ are seen and the first of these (near MJD 55100) occurs at a time of particularly low cadence (as low as 1 observation per 100 days). However there is an increase in power coincident with the major peak at MJD 56600, consistent with the power excesses in the AFB data.  As the difference between the profile in the two $\dot{\nu}$ states is too subtle to be seen by comparing profiles from individual observations, we average together all AFB profiles for which $\dot{\nu} < -1.740 \times 10^{-15}$ Hz s\textsuperscript{-1} to represent the weak spin-down state and $\dot{\nu} > -1.755 \times 10^{-15}$ Hz s\textsuperscript{-1} to represent the strong spin-down state.  The resulting profiles are shown in Figure \ref{srccgroup}(C) (panel ii) where there is a slight trailing edge excess in power when the pulsar is spinning-down most slowly (blue profile).  Conversely, there is a deficit in power in this phase region when the pulsar is spinning down more rapidly (red profile). These profiles bear strong similarity to those noted in LHK using the same AFB data. The strength of the correlation between $\dot{\nu}$ and the profile shape is also reflected in the correlation map (Figure \ref{srccgroup}(C)). The correlation is strongest at zero lag near profile phase 0.033 with recurrent strong peaks at intervals of 1500 days.

%% file: sections/results/1642.tex
\subsubsection{PSR B1642-03} \label{1642}

PSR B1642$-$03 is an older pulsar with a spin-period of 388 ms. While \cite{ran90} classified this pulsar as exhibiting a single component profile, \cite{kra94} noted weak outer conal components close to either side of the main pulse which become more dominant at higher frequencies.  The timing residuals show a quasi-periodic structure with sharp decreases towards negative values followed by a more gradual rise (see Figure \ref{resids}).  The radii of curvature of the peaks are smaller than those of the troughs; a common feature in many pulsars whose residuals show strong timing noise.  The evolution of the spin-down rate is shown in Figure \ref{nudots_prof1} and is characterised by stong narrow peaks on average every 1800-2000 days, though we note that the occurrence of peaks is distinctly less regular than in some other sources (e.g., PSR B1540$-$06). For example, a peak occurs at MJD 50000 followed by a subsequent peak $\sim$1400 days later. A further 2600 days elapses before $\dot{\nu}$ peaks again. The lack of regular periodicity is reflected in the Lomb-Scarge spectrum (Figure \ref{LS}) where multiple peaks are clustered around a frequency of 0.2 yr$\textsuperscript{-1}$ corresponding to a periodicity of 5 years ($\sim$1800 days).   The value of $\dot{\nu}$ oscillates about the mean with a peak-to-peak fractional amplitude $\Delta \dot{\nu} /  \dot{\nu} \approx 2.5$ per cent. The form of the variation is anti-symmetric about the peaks (where the pulsar is spinning down less rapidly) with a sharp decrease towards stronger spin-down followed by a more gradual, stochastic rise towards a weaker spin-down rate. 

\begin{figure*}
    \includegraphics[width=2.0\columnwidth]{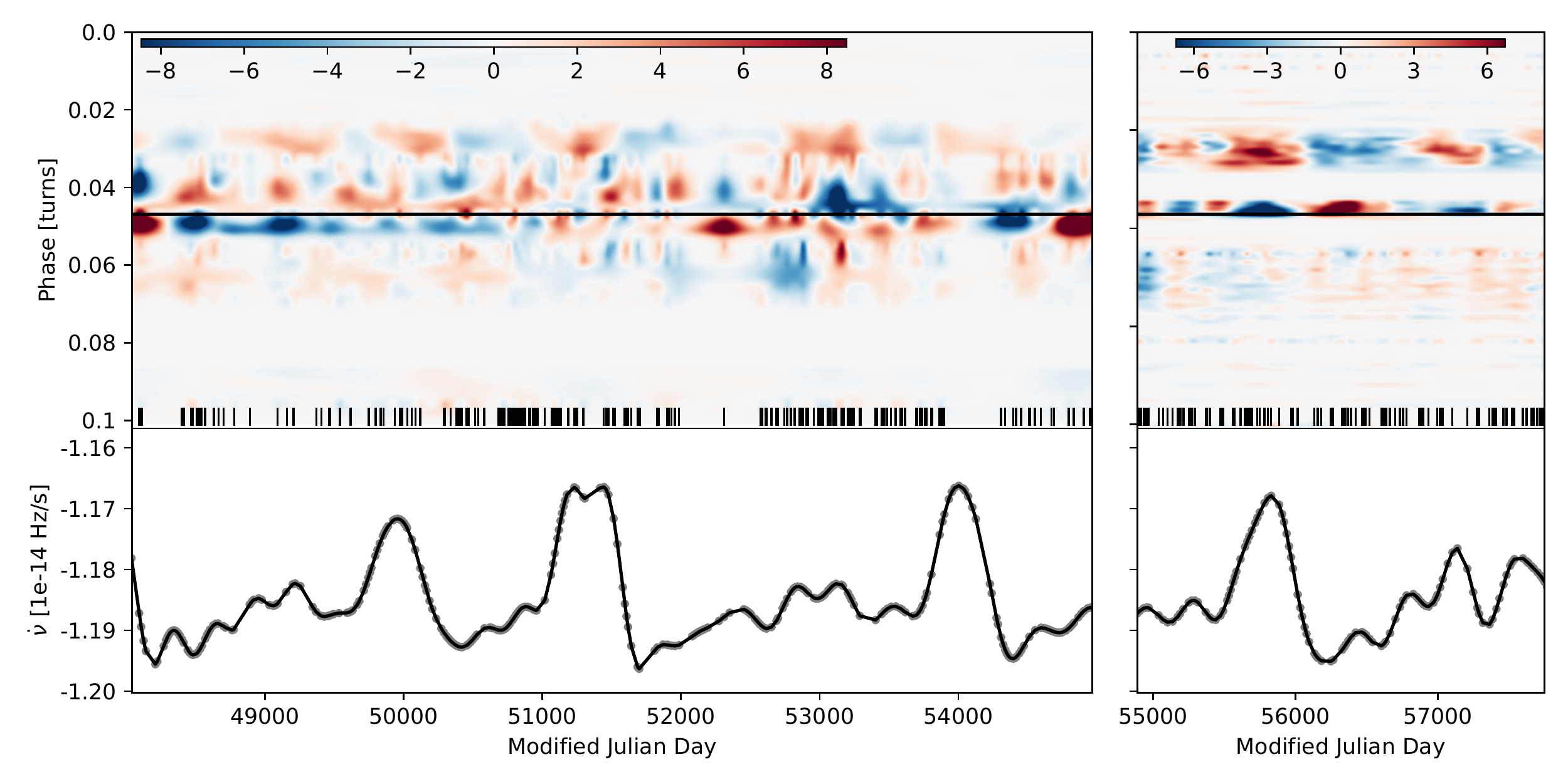} \\ 
    \caption[Profile and $\dot{\nu}$ variations in PSR B1642$-$03]{Profile variation maps and $\dot{\nu}$ variations in PSR B1642$-$03. As Figure \ref{0740_maps} otherwise.}
    \label{1642_map}
\end{figure*}

Figure \ref{1642_map} shows variablility in the shape of the pulse profile with coincident variations in the value of $\dot{\nu}$. This is particularly clear in the higher time-resolution DFB data.  Where the rate of spin-down is weakest (at the peaks in the lower panel of Figure \ref{1642_map}), there is a clear excess in power at the leading conal component of the pulse profile (phase $\sim$0.03). Conversely, where spin-down is strongest, this component is minimised. We also note that the relative amplitudes of the $\dot{\nu}$ values generally correlate with the amplitude of the strength of the leading component.  The most prominent peak in $\dot{\nu}$ near MJD 55800 is concident with a relatively strong leading component when compared to the following local maximum in $\dot{\nu}$ near MJD 57200 which is comparatively weak, although the observing cadence was notably lower in the latter. The correlation between $\dot{\nu}$ and the profile shape is substantially less clear in the AFB data, due to a large number of low S/N profiles and a highly variable cadence. However, subtle excesses are visible in the leading component when the pulsar is spinning down most weakly.

Similarly to PSR B1540$-$06, the difference in the profiles in each $\dot{\nu}$ state are not conspicuous from comparing profiles from single observations so we computed the median power in each phase bin, of all profiles observed in the low spin-down state (between MJDs 55613 and 55844) and compared this profile to the template profile, formed from averaging over all DFB epochs.  We repeated this for the high spin-down state (between MJDs 56155 and 56494, see Figure \ref{srccgroup}(D) (panel ii)). There is a clear difference in leading component power, relative to the template, when the pulsar is in one extreme spin-down state or the other.  At high/low spin-down (respectively blue/red lines) the leading component is $\sim$1 per cent weaker/stronger relative to the template.  The lower panel shows that the strongest correlation between $\dot{\nu}$ and the profile shape occurs at zero lag near phase $\sim$0.007. We also note variation in the profile near 0.027 phase (Figure \ref{srccgroup}(D)) in which a maximum in the correlation occurs between a lag of 750 and 1500 days where the large peak in $\dot{\nu}$ near MJD 55800 aligned with a some subtle variation in power between phases 0.05 and 0.07 in Figure \ref{1642_map}, though this trailing edge variability is less well correlated with $\dot{\nu}$.

%% file: sections/results/1822.tex
\subsubsection{PSR B1822$-$09} \label{1822}

PSR B1822$-$09 is one of the youngest pulsars in our sample with a characteristic age of $\sim$200 kyr. Mode-switching in PSR B1822$-$09 was first reported in \cite{fwm81}. Single pulse observations at multiple frequencies (e.g., \citealt{fwm81}; \citealt{gjk+94}) have shown that the pulsar spends the majority of its time in a bright (\emph{B}-) mode, characterised by a single component main-pulse profile (MP). For intervals of 5-10 minutes, the pulsar transitions to a quiet (\emph{Q}-)  mode in which the MP is accompanied by a leading component $\sim$35 ms ahead of the MP. Approximately 380 ms away (half of one period) from the MP, is an interpulse (IP) component which is weak or absent during the B-mode but sporadic in the Q-mode.

The timing residuals are dominated by two sharp turnover features nears MJDs 51000 and 52000. These (and three further features that are too small in amplitude to be seen in Figure \ref{resids}) have been described as \emph{slow-glitches}, characterised by a permanent increase in the spin-frequency, with no associated change to the spin-down rate (\citealt{zww+04}; \citealt{sha07}). However, LHK noted that these features were better explained by the pulsar entering a weaker spin-down state for short periods of time.  The $\dot{\nu}$ variations are shown in Figure \ref{nudots_prof1} with the reported glitches (slow or otherwise) clearly visible as peaks in the time series. LHK noted that these weak spin-down states were associated with the pulsar spending a larger fraction of time in the B-mode during these times. When the pulsar is in the otherwise stable high-$\dot{\nu}$ state, mode-switching between the two extreme profile shapes occurs rapidly on timescales of 20 minutes or less.  We measure the fractional change in $\dot{\nu}$ associated with the MJD 51000 transition to be $\Delta \dot{\nu} \approx 2.7$ per cent - this is somewhat smaller than the 3.3 per cent recorded in LHK, though we note that our covariance lengthscale of 221 days (Table \ref{cov}) is approximately double the stride-averaging window that was used by LHK. We note that no glitches or transitions to a significantly weaker spin-down mode have occured in our data since the glitch of MJD 54114.

\begin{figure}
    \includegraphics[width=1.0\columnwidth]{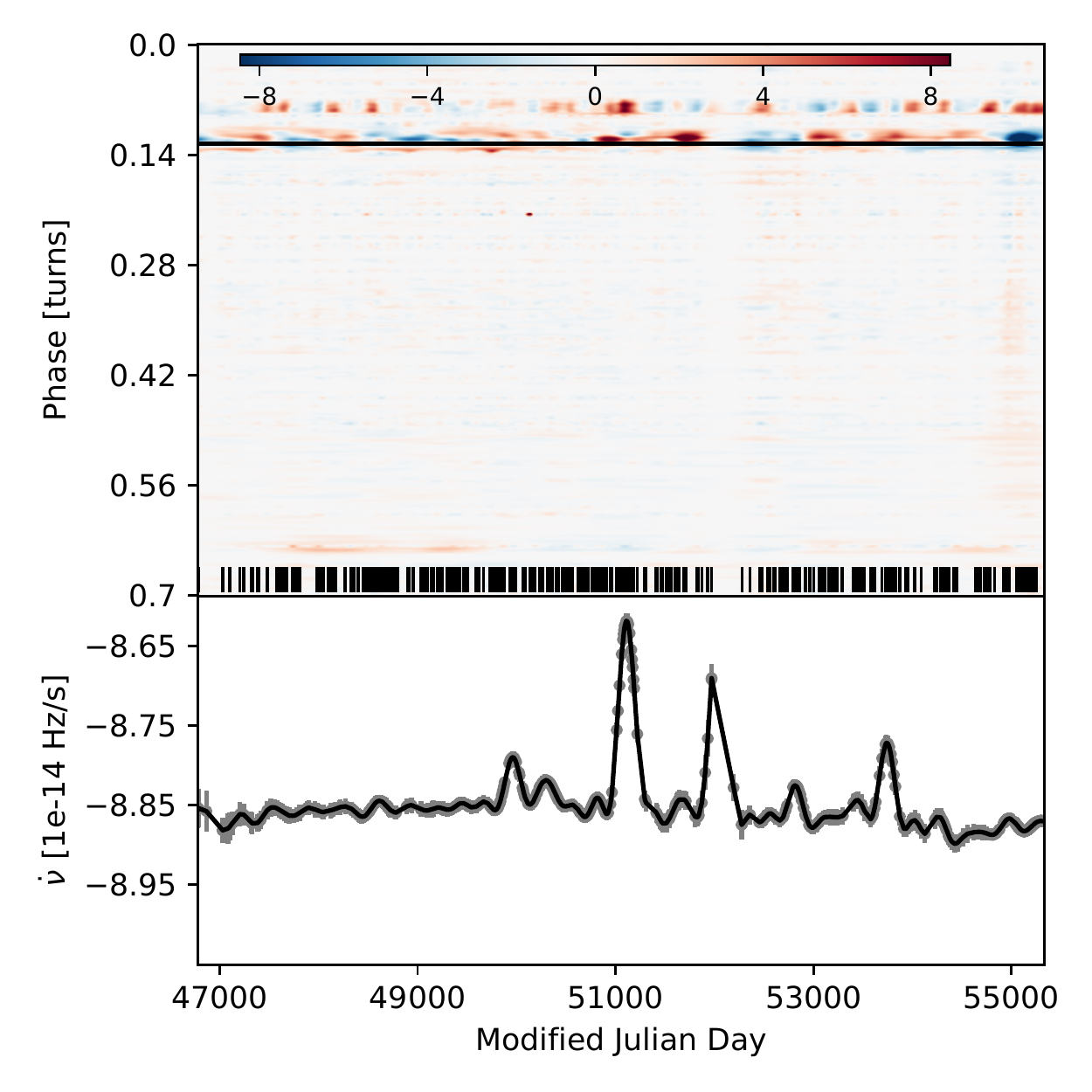}
    \caption[Profile and $\dot{\nu}$ variations in PSR B1822$-$09]{Profile variation maps and $\dot{\nu}$ variations in PSR B1822$-$09 AFB data.  As Figure \ref{0740_maps} otherwise.}
    \label{1822_map}
\end{figure}

Figure \ref{1822_map} shows the profile variability and $\dot{\nu}$ variations over the AFB data set during which the sharp peaks in $\dot{\nu}$ occured. The vertical axis on the variability map shows $\sim$70 per cent of one full turn of the pulsar with the precursor (PC), main pulse (MP) and interpulse (IP) components located near phases 0.09, 0.12 and 0.65 respectively. The PC was noted to be most active near the strong $\dot{\nu}$ spikes near MJDs 51000 and 52000 by LHK. We find this to be the case near MJD 51000 though the cadence near MJD 52000 is notably lower than that in LHK making the PC component here less clear. This is because a number of profiles observed around this epoch were not used in our analysis due to RFI contamination.  Similar brightening of the PC occurs at later epochs (particularly near MJDs 52500, 53250, 54000, 54800 and 55000) and appear to be correlated with smaller peaks in $\dot{\nu}$. We note however, that the somewhat larger $\dot{\nu}$ peaks near MJDs 52800 and 53800 do not correlate with an increase in PC flux. No significant changes are seen in the IP flux during this period, though its emission is considerably weaker and so small changes may be too subtle to be seen.  Figure \ref{srccgroup}(E) (panel ii) shows the differences between three profile morphologies in PSR B1822$-$09. The red (Q-mode) profile is associated with the strong spin-down state in which the PC component is absent and the IP is present. The blue profile (B-mode) shows the opposite case when the pulsar is in the weaker spin-down state (see \S \ref{disc}). The DFB data revealed no significant profile variations and so is not included here, though as mentioned above, no significant changes in $\dot{\nu}$ were seen after MJD 54114. 

%% file: sections/results/1828.tex
\subsubsection{PSR B1828$-$11} \label{1828}
PSR B1828$-$11 is the youngest pulsar in our sample, with a characteristic age of just 100 kyr. The variations in $\dot{\nu}$ and the profile shape in PSR B1828$-$11, as well as their correlation is extensively discussed in the literature, having first been reported in \cite{lss+00}. The oscillatory behaviour of the spin-down rate and the pulse shape has been associated with many phenomena, including the star freely precessing \citep{sls00}, precessing due to the presence of a fossil accretion disk \citep{qxxw03}, precessive torques exerted by an exotic companion \citep{lyx07} and free precession due to crustal strain \citep{jones12}. \cite{slk+18} reported correlated emission and spin-down changes in PSR B1828$-$11, showing that the pulsar exhibits two stables mode occuring over a cycle time of $\sim$500 days, and attributed the variations to magnetospheric switching. The pulse profile is characterised by two distinct shapes, one being a single component profile, and the other exhibiting much greater power at the leading edge (LHK) and BKJ revealed differences in flux density as well as profile shape, in the two spin-down states, characterised by the pulsar being brighter when the profile is single-peaked.  

The variation in $\dot{\nu}$ over time is shown in Figure \ref{nudots_prof1} and is remarkably periodic with sharp, high amplitude major peaks interrupted by lower amplitude minor peaks with a cycle time between major peaks of $\sim$500 days. The Lomb-Scargle spectrum (Figure \ref{LS}) shows a very high-Q\footnote{The Quality factor (or Q-factor) describes the power per unit width of a peak. High-Q suggests that spectral power is spread over a narrow range of frequencies.} peak at 0.75 yr\textsuperscript{-1} corresponding to 1.32 years (or $\sim$500 days) consistent with Figure \ref{nudots_prof1}. A second harmonically related peak occurs at 1.5 yr\textsuperscript{-1}. The difference in spin-down values between the major peaks corresponds to a peak-to-peak fractional amplitude of $\Delta \dot{\nu} / \dot{\nu} \approx 0.6$ per cent.  A slow linear rise in the overall value of $\dot{\nu}$ is seen across the dataset due to a non-zero $\ddot{\nu}$ which has not been included in our timing model.

\begin{figure*}
    \includegraphics[width=2.0\columnwidth]{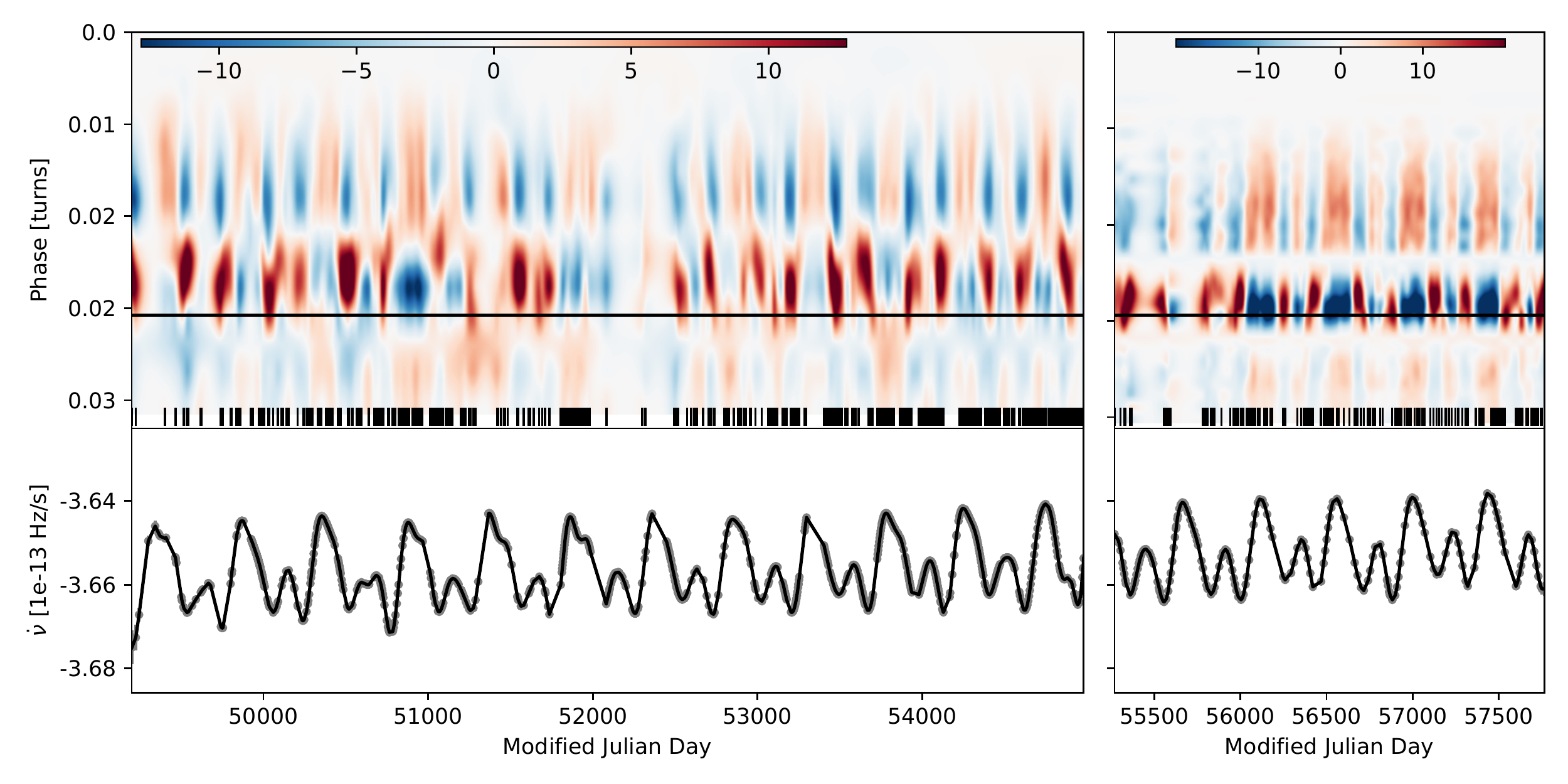} \\ 
    
    \caption[Profile and $\dot{\nu}$ variations in PSR B1828$-$11]{Profile variation maps and $\dot{\nu}$ variations in PSR B1828$-$11.  As Figure \ref{0740_maps} otherwise.}
    \label{1828_map}
\end{figure*}

The behaviour of the profile is shown in Figure \ref{1828_map} along with the coincident variations in $\dot{\nu}$. It is clear that PSR B1828$-$11 switches rapidly between two well defined emission states. In one state, there is a clear excess of power at the leading edge of the profile (phase $\sim$0.01) that is mirrored by a more subtle excess on the trailing edge.  In the other state, the pulse has a much narrower profile, characterised by the strong red regions of Figure \ref{1828_map} near the central peak. The relationship between the profile shape and $\dot{\nu}$ is striking with the magnitude of the power excess at the leading and trailing edges, very clearly tracing the major and minor peaks in the spin-down rate.

We show two examples of the pulse profile compared to the template in Figure \ref{srccgroup}(F). The blue trace shows a clear excess at the leading edge as well as the wider main component of the profile when the pulsar is spinning down more weakly.  The leading excess appears as a distinct component as seen in BKJ whereas in LHK it appears as a more gradual linear rise due to the lower time resolution of the data used in that study.  The red trace shows that the profile returns to a single-component state whilst the pulsar is in the strong spin-down mode, confirmed by BKJ to have a much greater flux density.  The lower panel of Figure \ref{srccgroup}(F), shows that the shape of the entire on-pulse region varies according to the value of $\dot{\nu}$. The correlation maps shows three distinct regions of correlation/anti-correlation corresponding to the leading edge, peak and trailing edge of the pulse.  The strongest correlation exists in the leading edge at a zero lag between the two time series. This corresponds to the low value of $|\dot{\nu}|$ being associated with an excess in leading edge power. The correlation is comparatively strong at a lag just short of 500 days, corresponding to the cycle time of the $\dot{\nu}$ variations.  Approximately half way between these two maxima is a region of weaker positive correlation associated with the major peaks of the $\dot{\nu}$ series and the minor peaks in the profile residuals.

%% file: sections/results/2035.tex
\subsubsection{PSR B2035+36}
Of the pulsars in LHK's sample, PSR B2035$+$36 exhibits the greatest peak-to-peak deflection in $\dot{\nu}$ in the form of a single large transition of $\Delta \dot{\nu} / \dot{\nu} = 13.28$ per cent. The profile has a triple-peaked structure comprising a central peak, preceded and followed by pre- and post-cursor wings. The emission switches between two very well defined shapes.  In the first the wings are bright and the triple-peakedness is clear.  In the second, the wings are of much lower amplitude (see Figure \ref{srccgroup}(G)).

We clearly resolve a single large transition in $\dot{\nu}$ showing that the pulsar spends most of its time in one of two extreme states (see Figure \ref{nudots_prof1}). By measuring the average value of $\dot{\nu}$ in each extreme state we find the transition amplitude $\Delta \dot{\nu} / \dot{\nu} \sim 12$ per cent.  This is somewhat less than the LHK value which can be understood from the slight linear rise in $\dot{\nu}$ that is only apparent with the longer dataset used here.  In addition to the single large switch, low amplitude secondary variations are seen within each of these extreme states with an apparent cycle time of $\sim 250$ days. In the earlier, weaker spin-down state, the peak-to-peak fractional amplitude $\Delta \dot{\nu} / \dot{\nu} \sim 2.4$ per cent.  

\begin{figure*}
    
    \includegraphics[width=2.0\columnwidth]{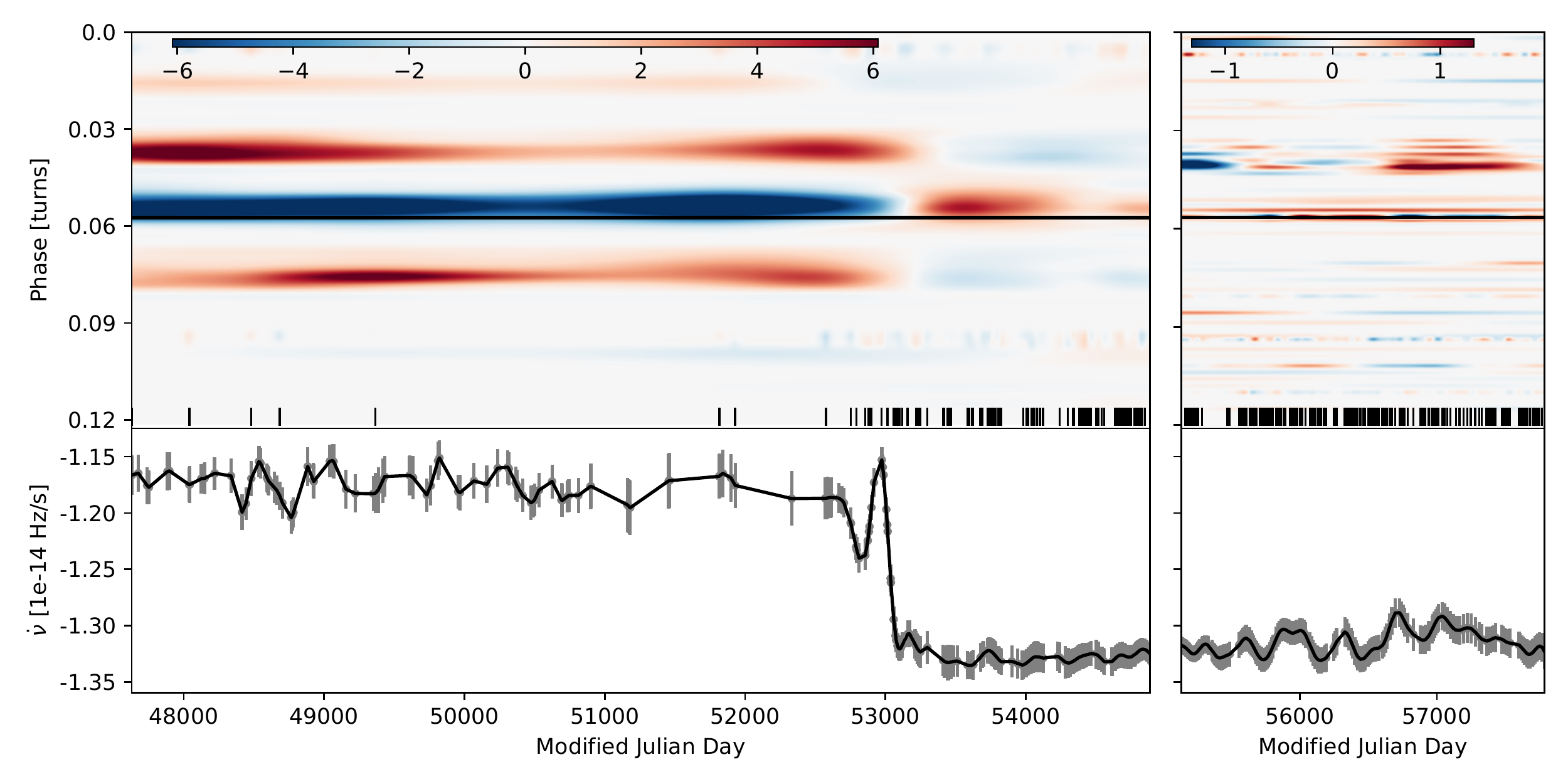} \\ 
    \caption[Profile and $\dot{\nu}$ variations in PSR B2035$+$36]{Profile variation maps and $\dot{\nu}$ variations in PSR B2035$+$36. As Figure \ref{0740_maps} otherwise.}
    \label{2035_map}
\end{figure*}

The large transition in $\dot{\nu}$ was accompanied by a clear change in the profile shape.  Figure \ref{2035_map} (upper panels) shows that coincident with the large transition, the wings of the profile dramatically reduced in amplitude, accompanied by a relative rise in the central peak power. It is not clear whether or not the secondary $\dot{\nu}$ modulations prior to the transition were accompanied by profile variations as there are too few observations obtained at that time.  In the DFB data, the pulsar spends all its time in only one of the extreme $\dot{\nu}$ states (corresponding to stronger spin-down). Though some slight variation is seen in the profile shape, it is of low amplitude and does not appear to correspond to the pulsar's $\dot{\nu}$ variability. 

To show the difference in the average pulse profile in each of the extreme spin-down states, we average together all of the AFB pulse profiles observed in each state (Figure \ref{srccgroup}(G)).  The drops in amplitude of the wing components between one state and the other are clear. The fact that the profile observed in the stronger spin-down state (red) and template (black dashes) traces are comparable is attributable to their being such a small number of observations in the low spin-down state (blue) mode, resulting in the template being biased towards that observed in the stronger spin-down state. The correlation between $\dot{\nu}$ and the pulse profile shape is shown in the lower panel of Figure \ref{srccgroup}(G).  The correlation/anti-correlation behaviour is clearly split over three distinct regions corresponding to the three profile components. The strongest correlation is at zero lag at phases corresponding to the leading and trailing edge emission. Conversely the strongest anti-correlation occurs at lag zero at a phase corresponding to the pulse peak.  As we have observed only a single large transition in both $\dot{\nu}$ and the profile shape, the correlation becomes slowly weaker as the magnitude of the lag increases.

%% file: sections/results/2043.tex
\subsubsection{PSR J2043$+$2740}

PSR J2043$+$2740 is the shortest period pulsar in our sample, completing one rotation every 96 ms. In the earlist JBO data (prior to MJD $\sim$52500), the profile comprised a sharp rise in power followed by a more gradual decay. The pulsar was noted by LHK to have undergone two large changes in $\dot{\nu}$ in a $\sim$10 year-long dataset.  The first of these changes, occurring around MJD 52500, was a large increase ($\sim$ 6 per cent) in the magnitude of $\dot{\nu}$. Coincident with this change was the appearance of a large second component at the trailing edge of the pulse.  The pulsar remained in this state for $\sim$3 years after which $\dot{\nu}$ returned towards its previous value along with a reduction in trailing edge power in the profile. Our analysis shows that J2043+2740 has exhibited a second example of this behaviour in the time since LHK, undergoing a similarly large increase in $\dot{\nu}$ around MJD 56500 (Figure \ref{nudots_prof1}), followed by a reverse transition of similar magnitude $\sim$3 years later around MJD 57500. The Lomb-Scargle periodogram for the PSR J2043$+$2740 shows a clear peak at a frequency of 0.1 yr\textsuperscript{-1}, corresponding to a cycle time of 10 years.

\begin{figure*}
    
    \includegraphics[width=2.0\columnwidth]{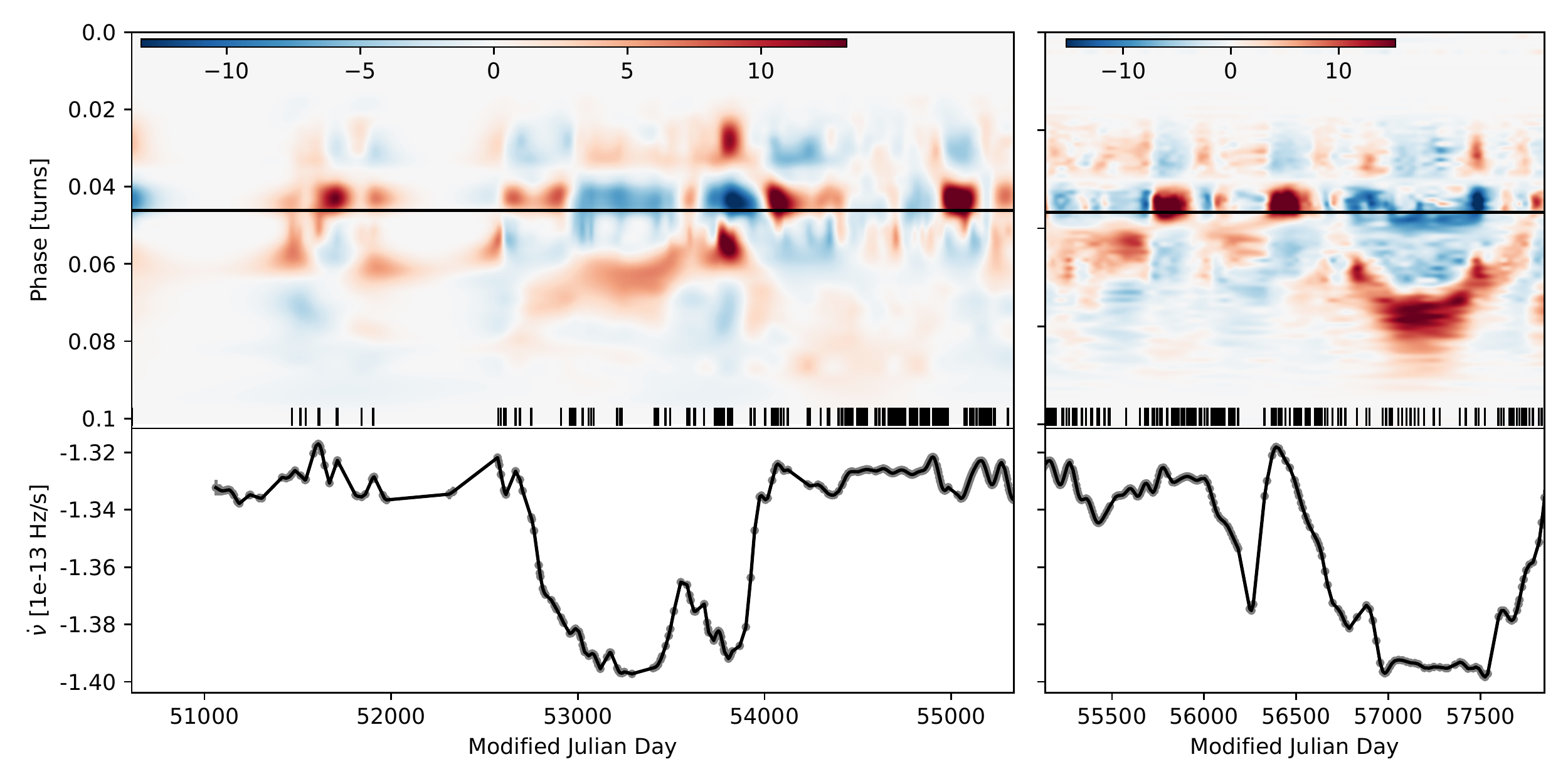} \\
    \caption[Profile and $\dot{\nu}$ variations in PSR J2043$+$2740]{Profile variation maps and $\dot{\nu}$ variations in PSR J2043$+$2740.  As Figure \ref{0740_maps} otherwise.}
    \label{2043_map}
\end{figure*}

Our variability maps (Figure \ref{2043_map}) confirm the increase in power of the trailing component of the pulse profile when the pulsar is spinning down more rapidly. We also note the occurrence of a strong $\dot{\nu}$ decrease that occured whilst the pulsar was in the, otherwise stable, high $|\dot{\nu}|$ state (near MJD 53750). Though observations were particularly sparse near this time, the variability map shows a slight rise, and then fall, in leading edge power coincident with this variation.

The two distinct pulse profiles of PSR J2043$+$2740 are shown in Figure \ref{srccgroup}(H) (panel ii). The red trace representing the pulse profile when $|\dot{\nu}|$ was high was formed from the addition of profiles observed between $56602 < \mathrm{MJD} < 57848$. Conversely, the blue trace representing the pulse profile when $|\dot{\nu}|$ was low was formed from all profiles observed between $55136 < \mathrm{MJD} < 55985$. The fact that the blue trace very closely resembles the template profile that was formed from all DFB profiles is because a larger number of observations were made of J2043$+$2740 in the low $|\dot{\nu}|$ state.  We note that the strongest correlation occurs at a lag of $\sim$ +400 days and the reason can be seen by inpection of the variability map in which the value of $\dot{\nu}$ begins its downward deflection approximately 400 days prior to the onset of the profile change.

In each of the (otherwise stable) emission states, the pulsar appears to undergo briefer transitions in which it assumes a value of $\dot{\nu}$ closer to that exhibited in the opposite state. For example, in the strong $\dot{\nu}$ state beginning at MJD 52750, $\dot{\nu}$ becomes briefly weaker $\sim$400 days before entering the main weak state at MJD 54000 and this is accompanied by an increase in the profile's central component power.  More dramatically, prior to entering the strong $\dot{\nu}$ near MJD 56750, $\dot{\nu}$ undergoes a brief period of strong spin-down near MJD 56250, though in this case, whether or not the emission correspondingly changes is less clear to due a lack of profile data near that time. 

\begin{figure*}
    \vspace{-2.5mm}
    \includegraphics[width=\columnwidth]{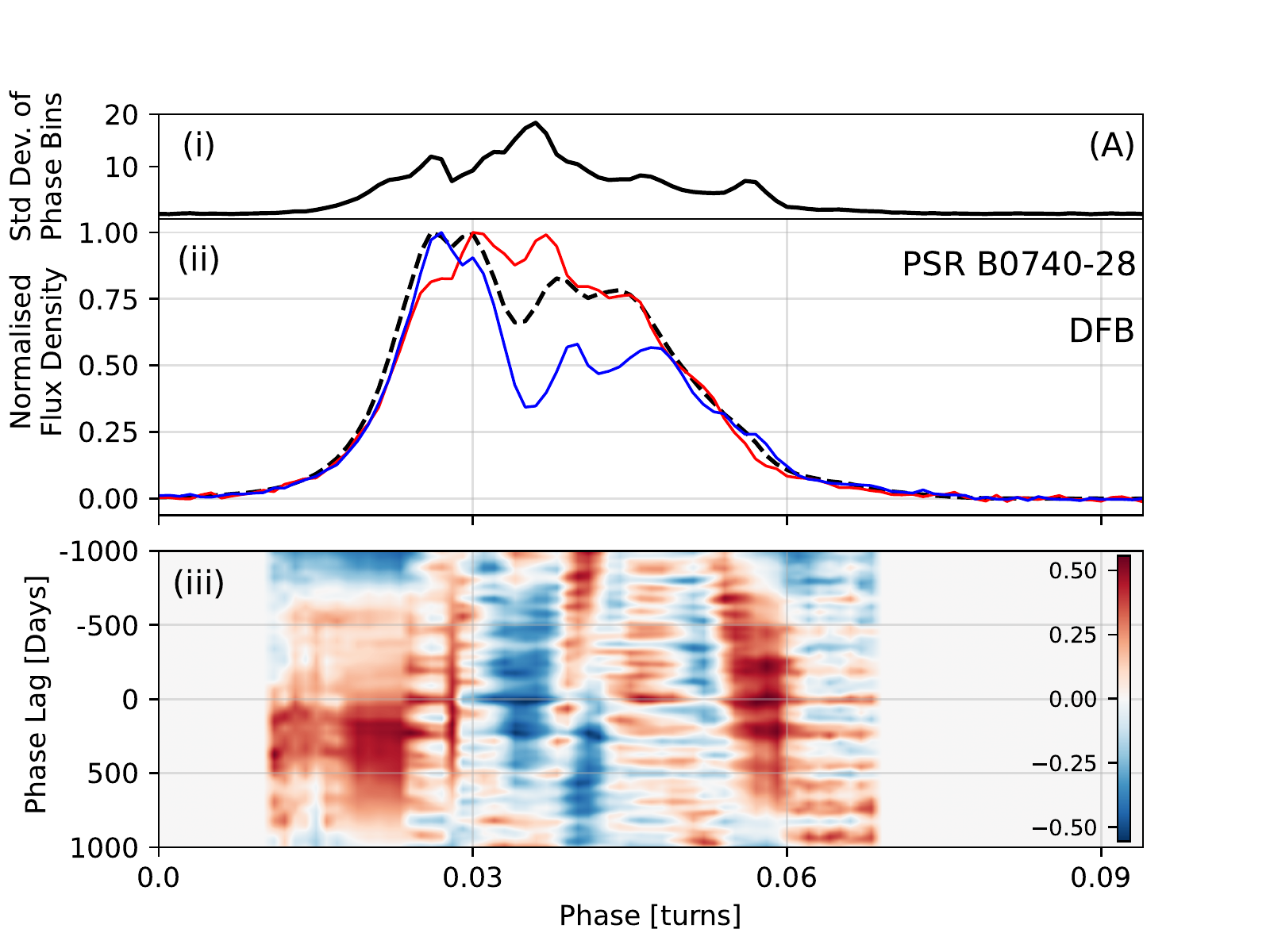}
    \includegraphics[width=\columnwidth]{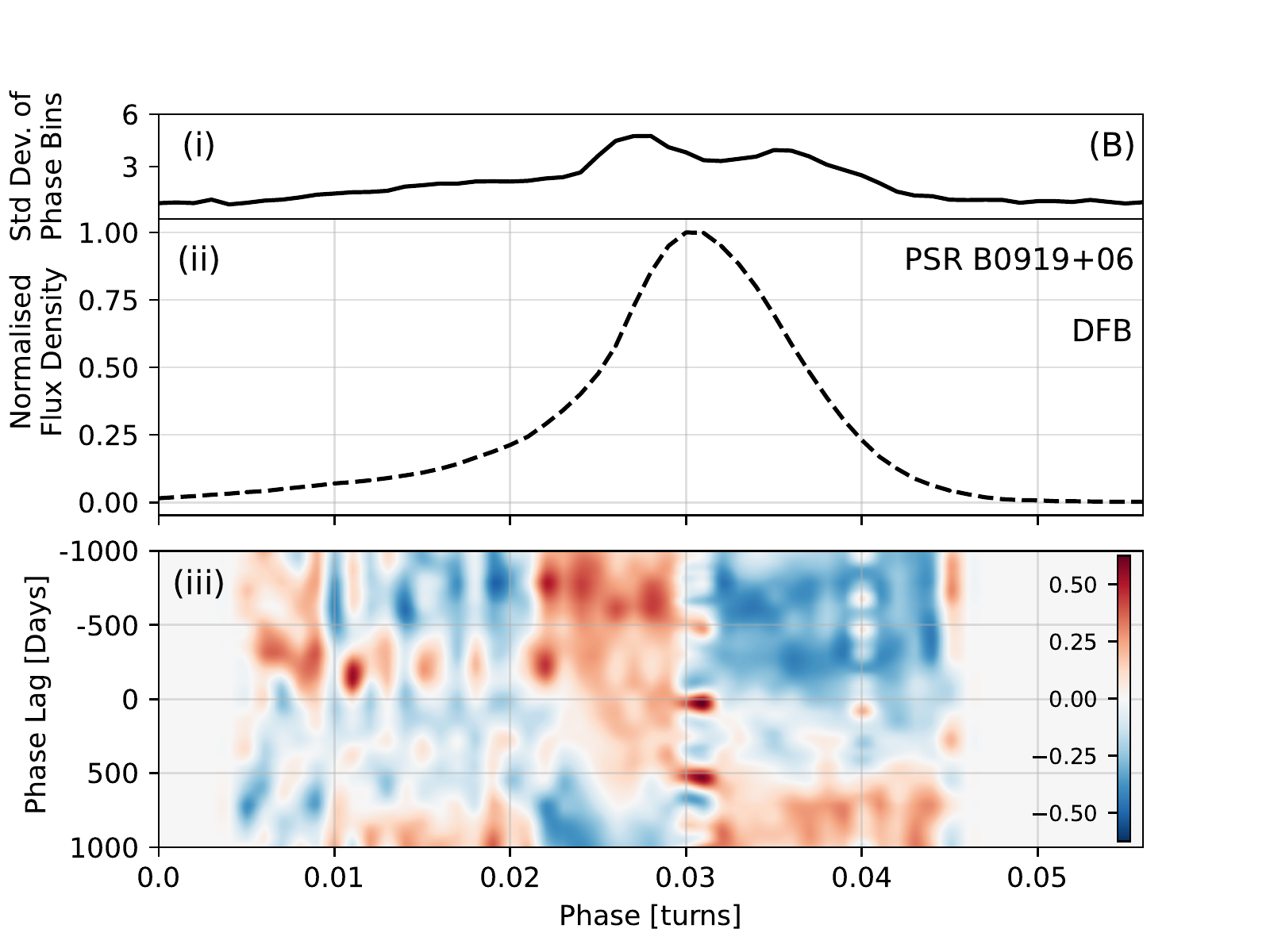} \\
    \vspace{-2.1mm}
    \includegraphics[width=\columnwidth]{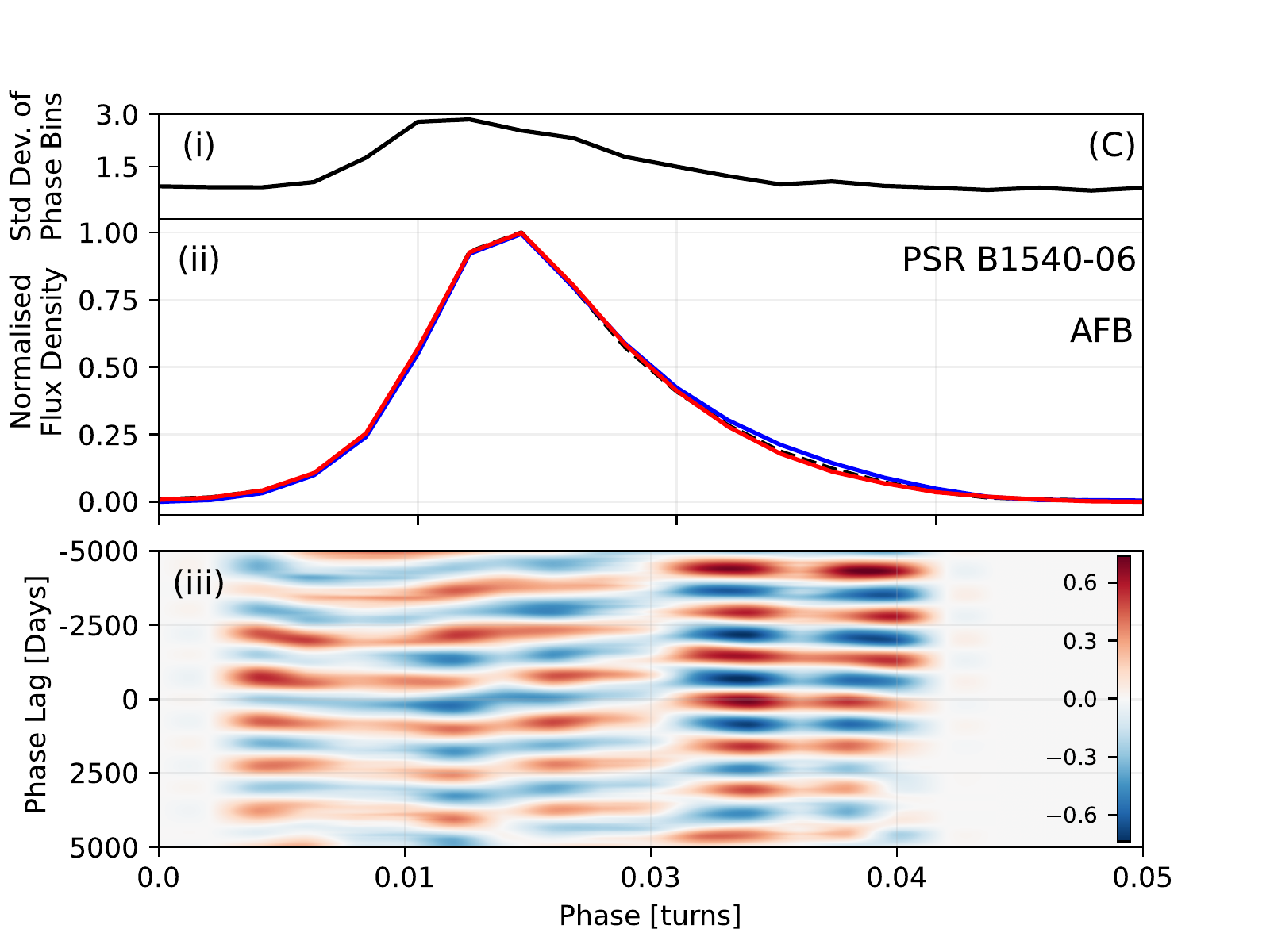} 
    \includegraphics[width=\columnwidth]{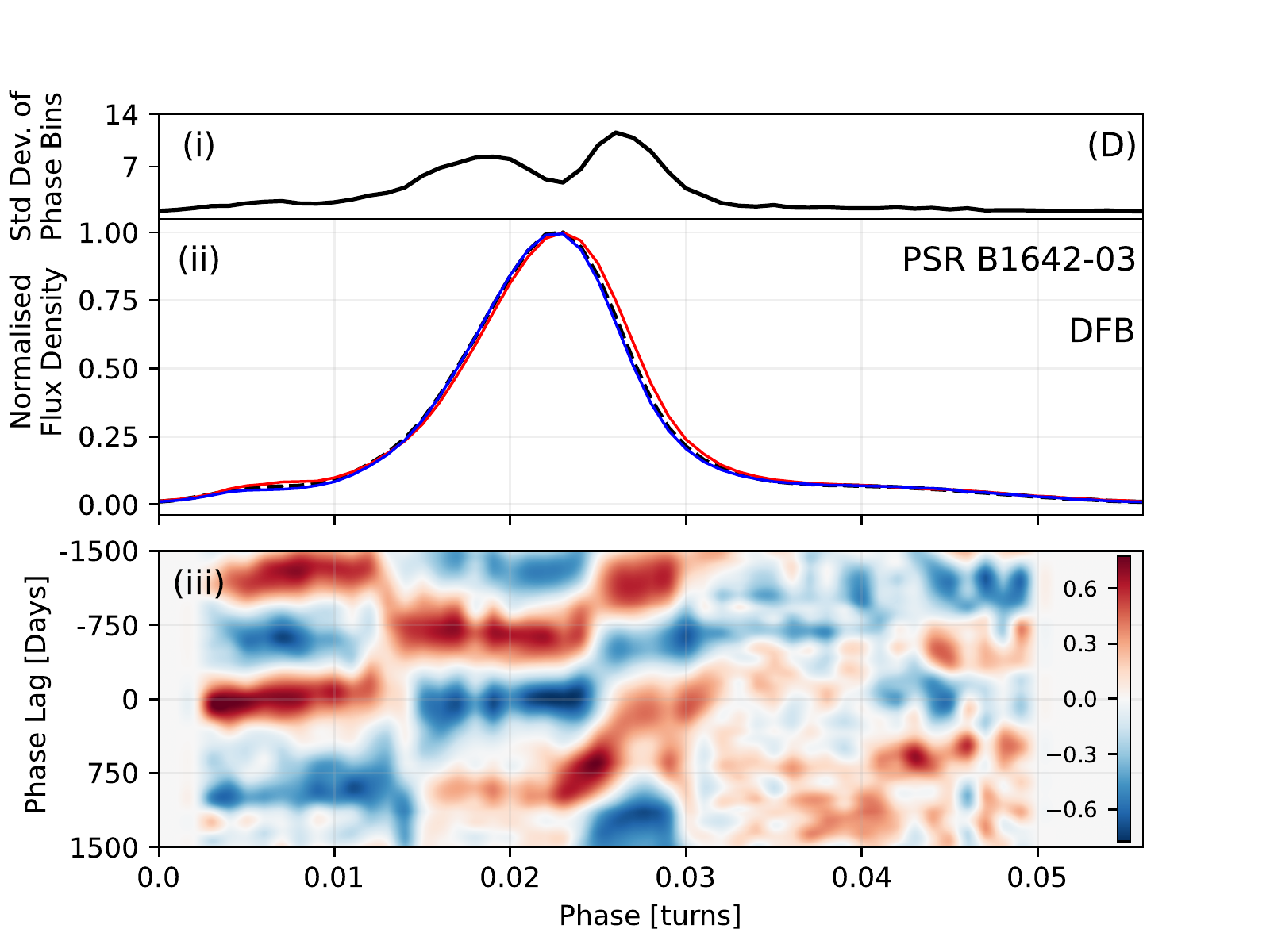} \\
    \vspace{-2.1mm}
    \includegraphics[width=\columnwidth]{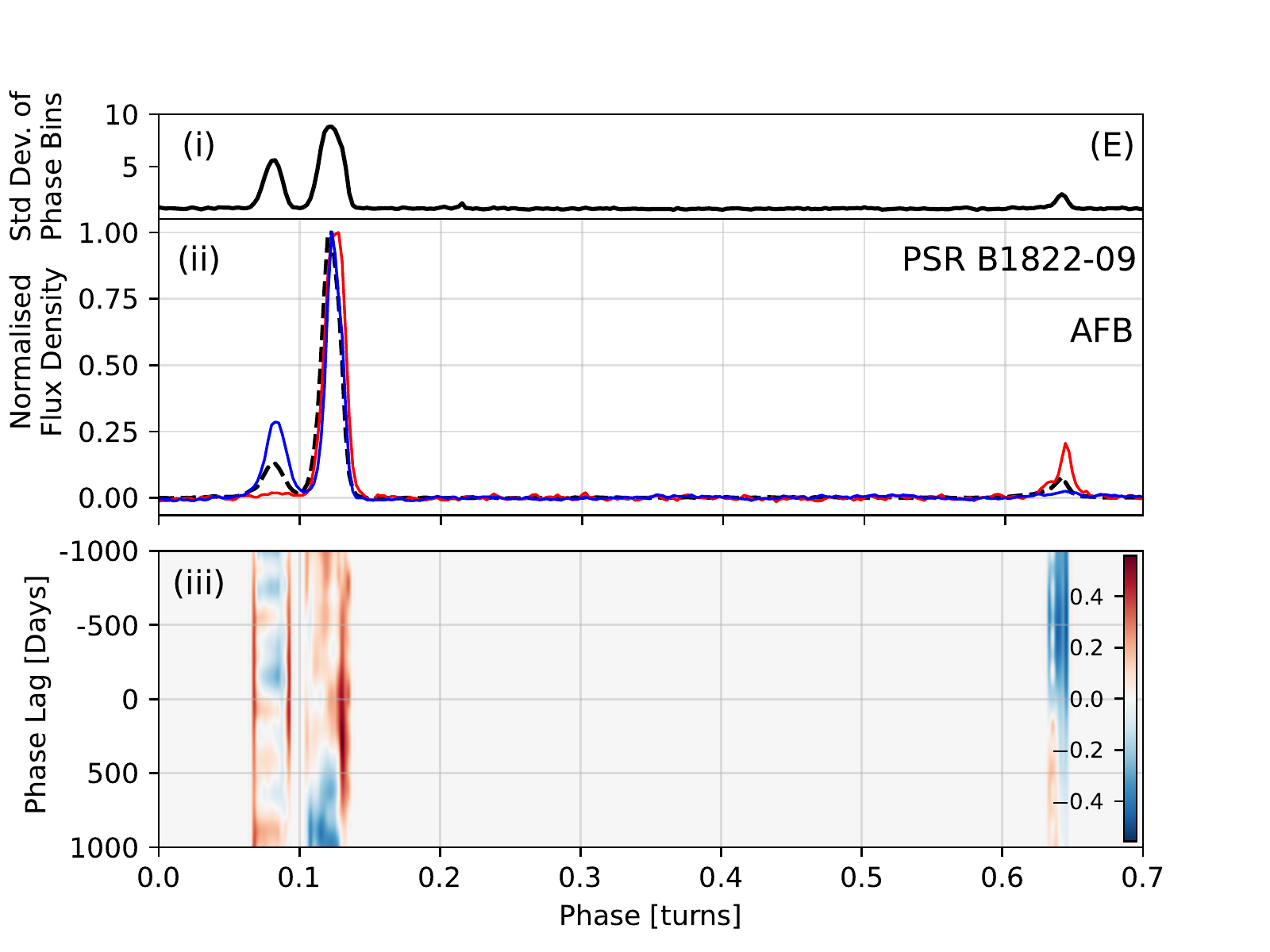}
    \includegraphics[width=\columnwidth]{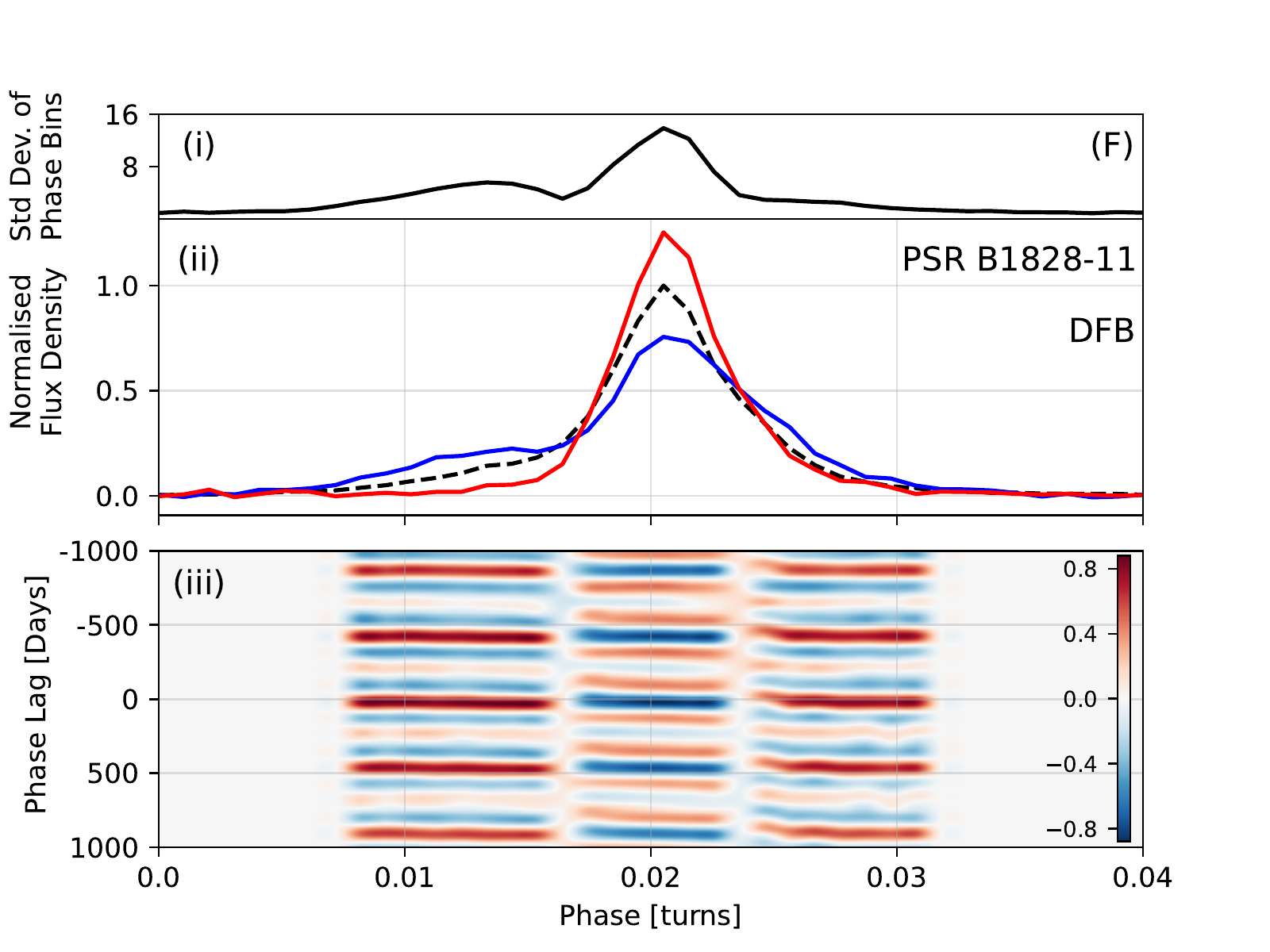}
    \caption{Pulse profiles and pulse shape/spin-down corrleation maps for the 8 pulsars in which we observe correlated emission and spin-down variability. The upper panels (i) show the standard deviations of each phase bin of the median profile in units of the off-pulse mean. Panels (ii) show the median pulse profiles (dashed lines) aggregated from all epochs in either the DFB or AFB datasets. We also show examples of the average pulse profiles observed when the pulsar is in a strong (red) and weak (blue) spin-down state. We do not include example profiles from PSR B0919+06 as the differences in shape between the two states are too slight.  In panels (iii) we show the Spearmann-Rank correlation coefficient (SRCC) for either the DFB or AFB dataset (chosen on the basis of the most conspicuous correlation), for $\dot{\nu}$ and pulse shape variations calculated for each unit of pulse phase. Red/blue regions respectively denote correlation/anti-correlation between the profile residuals for each phase bin, and the $\dot{\nu}$ time series. SRCC values are shown in the colour bars.}
    \label{srccgroup}
\end{figure*}

\begin{figure}
    \ContinuedFloat
    \vspace{-2.1mm}
    \includegraphics[width=\columnwidth]{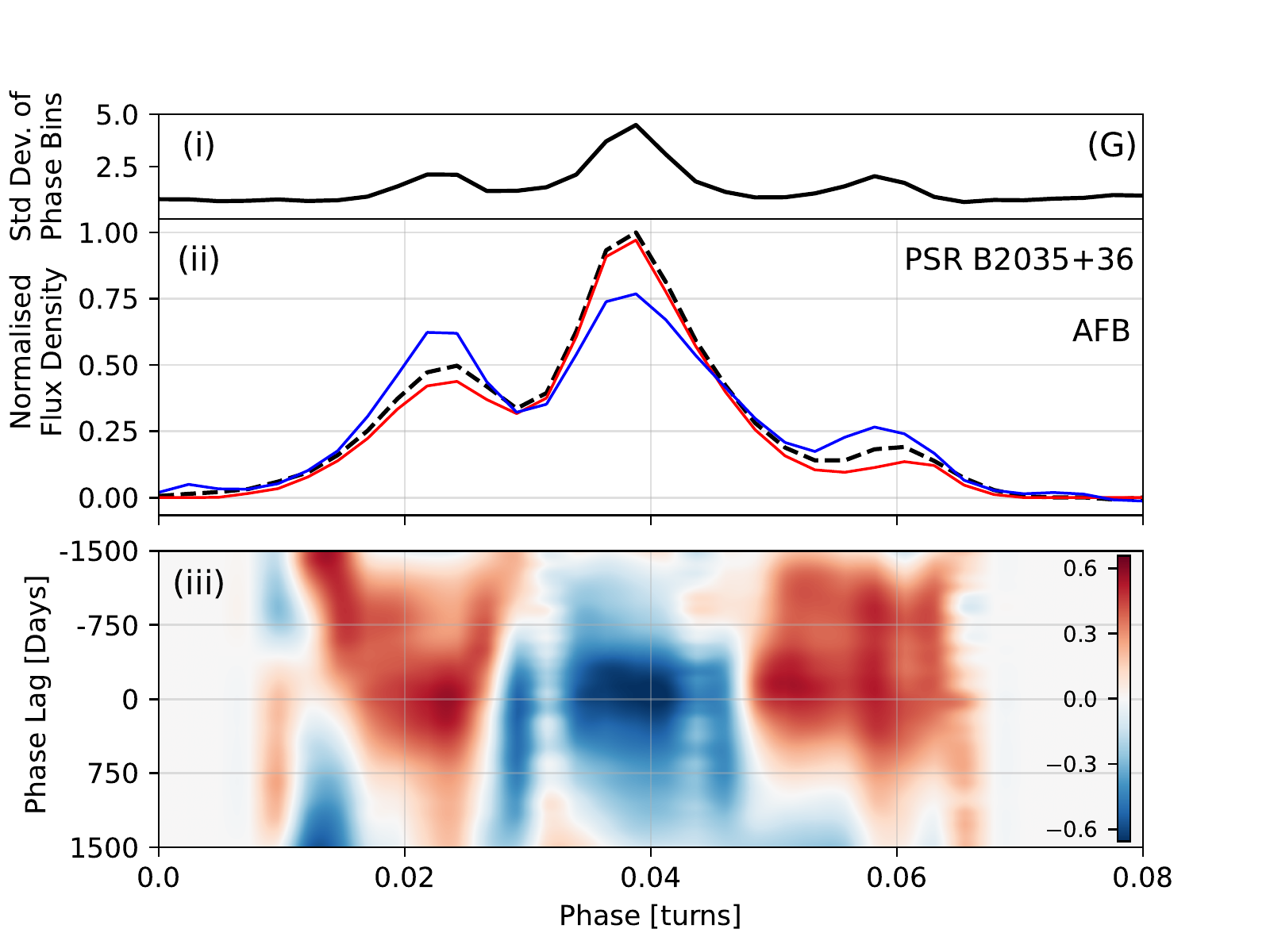} \\
    \includegraphics[width=\columnwidth]{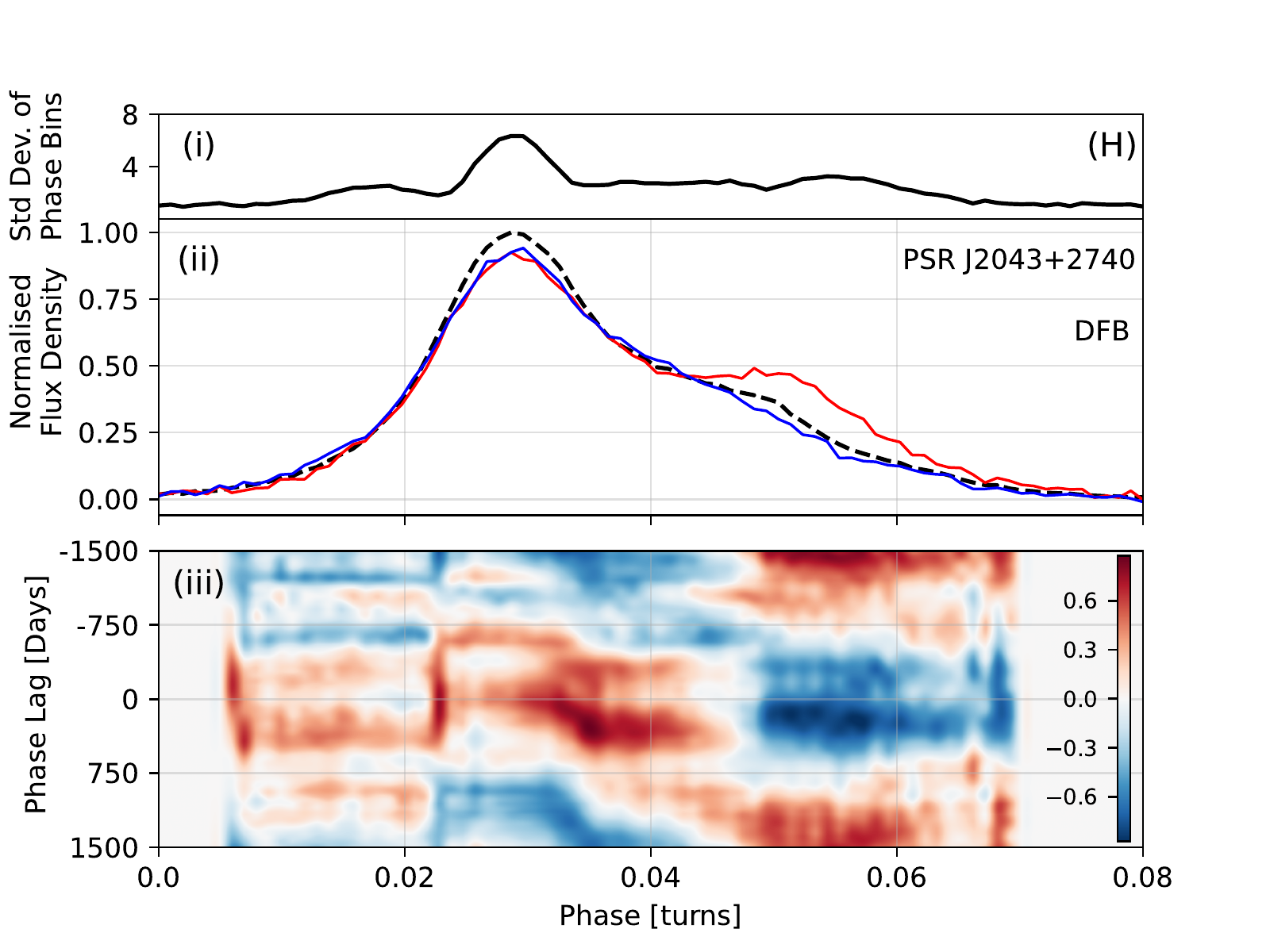}
    \caption[]{(continued)}    
\end{figure}

%% file: sections/results/noprofvar.tex
\subsection{Pulsars with no detected emission-rotation correlation}

\begin{figure}
    \includegraphics[width=\columnwidth]{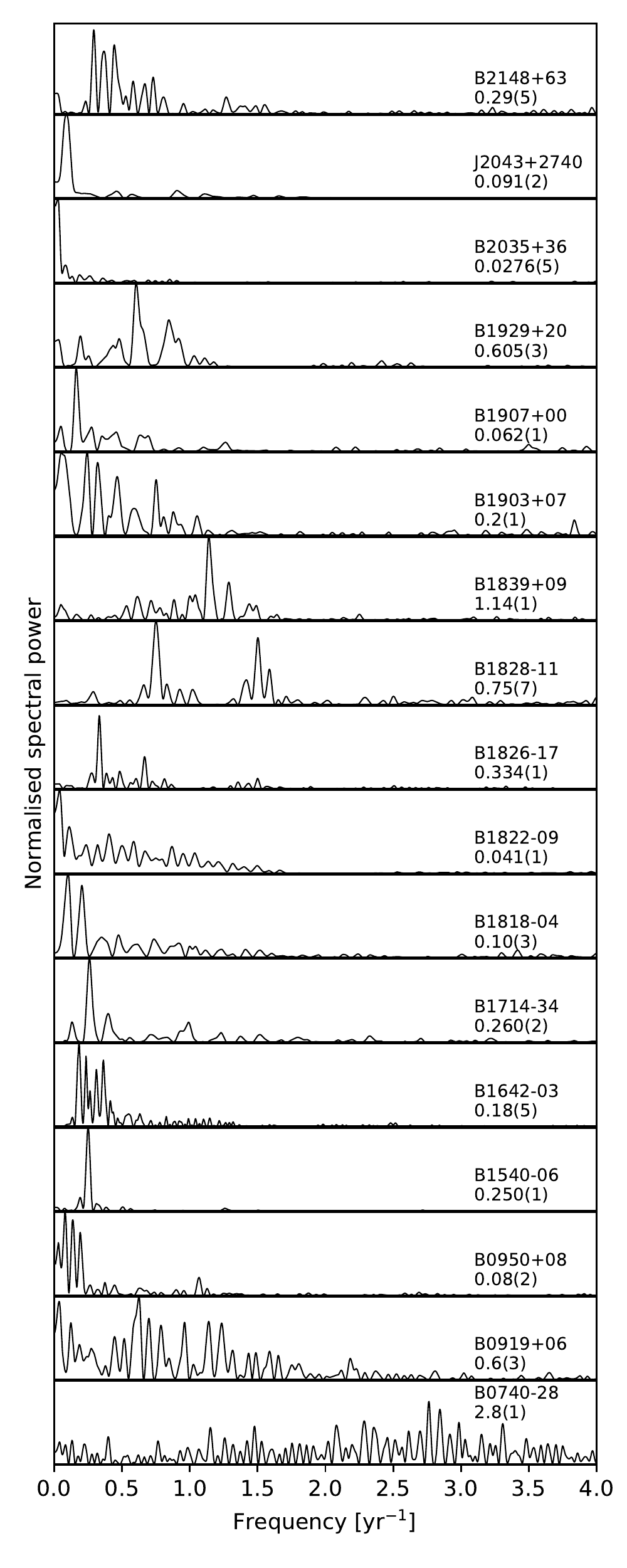}
    \caption[Lomb-Scargle spectra of the $\dot{\nu}$ evolution of 17 pulsars in a variability study]{Lomb-Scargle periodicity spectra of the $\dot{\nu}$ evolution for each pulsar in Figure \ref{nudots_prof1}. The values beneath the pulsar names are the frequencies in yr\textsuperscript{-1} corresponding to the maxima in the spectra.}
    \label{LS}
\end{figure}

\subsubsection{PSR B0950$+$08}

The timing residuals of PSR B0950$+$08 show no clear periodicity (Figure \ref{resids}) although we note that the local maxima are sharper than the local minima. Figure \ref{nudots_prof1} shows that the $\dot{\nu}$ evolution, though slowly varying, is more erratic than many of the pulsars studied here, conforming to no characterisic periodicities or peak-to-peak deflection amplitudes.  Strong spikes in $\dot{\nu}$ occur near MJDs, 44000 and 48000 with a $\Delta \dot{\nu} / \dot{\nu} \approx 0.8$ per cent.  A subsequent peak near MJD 54000 also appears but is comparatively weak. A later spike occurs near MJD 57000 when the pulsar assumes a lower $\dot{\nu}$ state for a longer period of time than the previous three. Our analysis did not identify any variations in the pulse profile of PSR B0950$+$08.


\subsubsection{PSR B1714$-$34}

The timing residuals for PSR B1714$-$34, shown in Figure \ref{resids}, show clear periodic variations, over a cycle time of $\sim$1500 days. There are no observations for 883 days between May 1994 (MJD 50044) and October 1996 (MJD 50927).  Analysis of the $\dot{\nu}$ variations (Figure \ref{nudots_prof1}) shows clear periodic behaviour which also cycles over $\sim$1500 days. The value of $\dot{\nu}$ exhibits sawtooth-like variations rising gradually towards lower values before sharply dropping towards higher values.  We compute a peak-to-peak fractional amplitude of $\Delta \dot{\nu} / \dot{\nu} \approx 0.9$ per cent, marginally higher than that reported in LHK (0.79 per cent).  Analysis of the spectral power of the $\dot{\nu}$ variations reveals a strong periodicity at 0.26(2) yr\textsuperscript{-1} corresponding to a cycle time of 1410 days. 

Our GP analysis revealed no evidence of profile variations. To search for subtle variations in the average profile in each $\dot{\nu}$ state, we formed two integrated profiles from all individual epochs where $\dot{\nu}$ was computed to be strongest ($\dot{\nu} < -2.277 \times 10^{-14}$ Hz s\textsuperscript{-1}) and weakest ($\dot{\nu} > -2.273 \times 10^{-14}$ Hz s\textsuperscript{-1}) respectively. We find no significant difference between the two integrated profiles.


\subsubsection{PSR B1818-04}

The timing residuals of PSR B1818$-$04 show clear oscillatory behaviour (Figure \ref{resids}), and were noted by \cite{hlk10} to exhibit a cycle time of 7-10 years. The variations in $\dot{\nu}$ (Figure \ref{nudots_prof1}) show that PSR B1818$-$04 spends extended periods of time in one or another spin-down state. Over the course of our dataset ($\sim$29 years) the pulsar has completed three intervals in the stronger and two in the weaker of these states.  The time spent in the stronger state is shorter ($\sim$1000 days) than that spent in the weak state ($\sim$2300 days). The pulsar transitions into the strong spin-down state approximately every 10 years and this is reflected in the maximum of the Lomb-Scargle spectral power (Figure \ref{LS}) which occurs at 0.1 yr\textsuperscript{-1}. The peak-to-peak fractional amplitude of these transitions is  $\Delta \dot{\nu} / \nu \approx 0.9$ per cent.  In addition to these \emph{major} transistions, there are minor oscillations within each of the otherwise stable modes which cycle with an approximate timescale of 420 days and a $\Delta \dot{\nu} / \nu \approx 0.3$ per cent. We note that the cadence in later (DFB) data (since MJD $\sim 54000$) is significantly lower that in earlier (AFB) data.

We find no significant shape variations in the pulse profile in either of the DFB or AFB datasets for this source. To further verify this we formed two average pulse profiles from the AFB data  (as this has the higher average cadence) - composed of individual epochs where $\dot{\nu}$ was computed to be strongest ($\dot{\nu} < -1.772 \times 10^{-14}$ Hz s\textsuperscript{-1}) and weakest ($\dot{\nu} > -1.770 \times 10^{-14}$ Hz s\textsuperscript{-1}). We find no significant differences between the two.

\subsubsection{PSR B1826$-$17}
The 26 years of timing residuals of PSR B1826$-$17, shown in Figure \ref{resids}, have a noticably smaller radius of curvature at the local maxima than at the local minima, consistent with findings by LHK and \cite{hlk10}.  The timing residuals also exhibit clear short-term quasi-periodicity, superimposed on a more long-term variation. \cite{hlk10} noted a significant peak in the power spectrum of the residuals at 2.9 years.  

\begin{figure}
    \includegraphics[width=1.0\columnwidth]{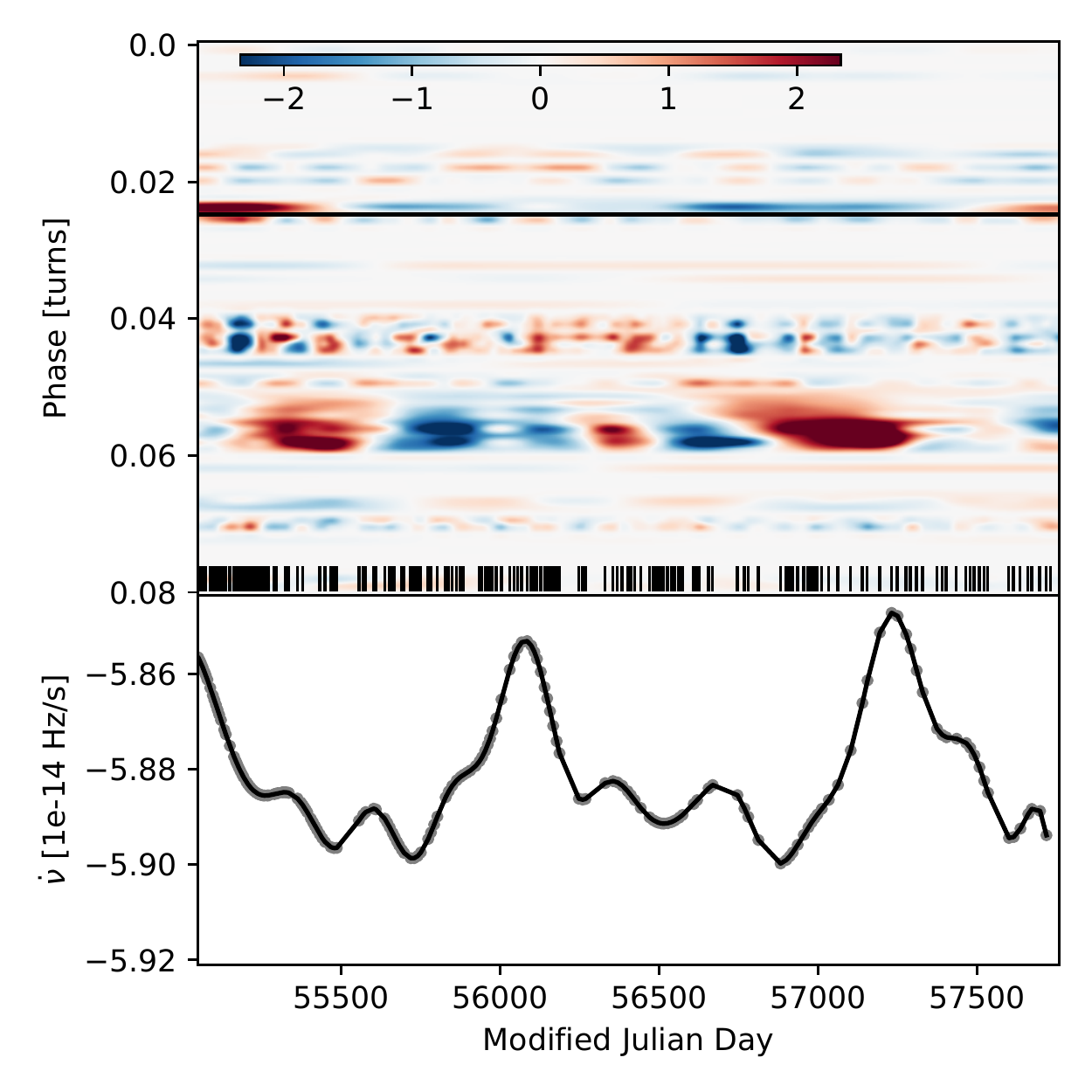}
    \caption[Profile and $\dot{\nu}$ variations in PSR B1826$-$17]{\label{maps_dfb_1826}Profile variation maps and $\dot{\nu}$ variations in PSR B1826$-$17. Only the DFB data is shown here. As Figure \ref{0740_maps} otherwise.}
    \label{1826_map}
\end{figure}

The computed spin-down variations are shown in Figure \ref{nudots_prof1} and show a periodic nature with major peaks occurring roughly every 1100 days. These peaks show a peak-to-peak fractional amplitude of $\Delta \dot{\nu} / \dot{\nu} \approx 0.7$ per cent. This was also noted by \cite{lyne13}.  Between major peaks are smaller, lower amplitude peaks reminsicent of those seen in PSRs B1828$-$11 and B0919$+$06, though in this case, the major and minor peaks are significantly less well defined. The modulations in earlier data were much noisier than more recently, because the cadence in the AFB dataset was much lower than in the DFB dataset.  A Lomb-Scargle spectral analysis (Figure \ref{LS}) shows a strong, narrow peak that corresponds to a periodicity in $\dot{\nu}$ of $\sim$1100 days. Figure \ref{1826_map} shows some subtle changes in the amplitude of the trailing component of the triple-peaked profile. \cite{lyne13} also noted a 10 per cent change in the ratio between the central and leading/trailing components of the profile.  These variations however do not appear correlated with the spin-down rate.

\subsubsection{PSR B1839$+$09}

The variations in $\dot{\nu}$ of PSR B1839$+$09 are shown in Figure \ref{nudots_prof1} and exhibit regular sharp oscillations ($\Delta \dot{\nu} / \nu \approx 2.7$ per cent) every $\sim$300 days. Lomb-Scargle spectral analysis of the $\dot{\nu}$ variations reveals a peak at 1.1 yr\textsuperscript{-1}.  Over the $\sim$2546 days covered in the DFB dataset, just 40 observations were included in our analysis.  Our analysis reveals no significant profile shape variations in either the DFB or AFB datasets.  To search for more subtle average variations we combined AFB profiles observed when $\dot{\nu} > -7.45 \times 10^{-15}$ Hz s\textsuperscript{-1} and when $\dot{\nu} < -7.55 \times 10^{-15}$ Hz s\textsuperscript{-1}.  We find no detectable differences between the two.

\subsubsection{PSR B1903$+$07}

Whilst the variations in $\dot{\nu}$ are approximately periodic, the amplitudes of the variations are highly irregular across the dataset (Figure \ref{nudots_prof1}).  Prior to MJD 50000, $\dot{\nu}$ varies by a $\Delta \dot{\nu} / \dot{\nu} \sim 5.9$ per cent over a cycle time of $\sim 1000$ days. At $50000 \leq \mathrm{MJD} \leq  53000$, $\dot{\nu}$ cycles by $\Delta \dot{\nu} / \dot{\nu} \sim 3.4$ per cent over a timescale of $\sim 500$ days and is consistenly weaker than the mean value in this range.  The overall value of $\Delta \dot{\nu} / \dot{\nu} \sim 6.8$ per cent.  

Our GP analysis does not reveal any significant profile variations in either the AFB or DFB datasets. We note however, that due to the low S/N and low cadence (typical cadence $\sim$38 days), the template profile generated for this source is noisy, thereby reducing our sensitivity to any variations. Combining all profiles in the high ($\dot{\nu} < -1.19 \times 10^{-14}$) and low ($\dot{\nu} > -1.17 \times 10^{-14}$) spin-down states in the AFB dataset respectively and comparing the resulting integrated profile, we find no clear differences between the two.

\subsubsection{PSR B1907$+$00}

PSR B1907$+$00 is the slowest pulsar in our sample with a spin-period of 1.02 s.  Timing irregularities in this source were noted by \cite{hlk+04}. The value of $\dot{\nu}$ oscillates about the mean with a peak-to-peak fractional amplitude of $\Delta \dot{\nu} / \dot{\nu} \sim 0.8$ per cent. The variations comprise consecutive major and minor peak with a cycle time between the major peaks of $\sim$ 6 years. LHK noted a peak in the power spectrum of the $\dot{\nu}$ variations at 0.15(2) yr\textsuperscript{-1} corresponding to a periodicity of 6.2 years. Our analysis reveals a peak at the same location (Figure \ref{LS}).  Our analysis reveals no significant profile shape variatons in either the DFB or AFB datasets.

\subsubsection{PSR B1929$+$20}

The timing residuals of PSR B1929$+$20 show clear periodic variations and this was also noted by \cite{hlk+04} and \cite{hlk10}. The value of $\dot{\nu}$ oscillates with a peak-to-peak fractional amplitude $\Delta \dot{\nu} / \dot{\nu} \approx 0.31$ per cent with an approximate periodicity of 1.7 years.  There is a notable reduction in $\Delta \dot{\nu} / \dot{\nu}$  between $51000 < \mathrm{MJD} < 53000$, though this may reflect the reduced cadence near that time.  Analysis of the periodicities of the $\dot{\nu}$ variations shows a strong peak around 0.6 yr\textsuperscript{-1}, (Figure \ref{LS}). We were unable to identify any pulse profile variations. Individual profiles for this pulsar in both the DFB and AFB datasets have a very low S/N. The low cadence with which this source is observed (just 75 observations since October 2009) makes it unfeasible to combined observations in order to search for subtle average pulse shape variations.


\subsubsection{PSR B2148$+$63}

The timing residuals for PSR B2148$+$63 show significiant periodic structure in which the local minima are sharper than the maxima, contrary to the case in other pulsars that exhibit timing noise such as PSRs B1642$-$03 and B1826$-$17. This was also noted by \cite{hlk10}.  Figure \ref{nudots_prof1} shows that the value of $\dot{\nu}$ oscillates about the mean with a peak-to-peak fractional amplitude of $\Delta \dot{\nu} / \dot{\nu} \approx 2.1$ per cent, somewhat greater than the value measured in LHK (1.7 per cent).  We attribute this to the fact that LHK used a stride window of 600 days which is considerably wider than the lengthscale optimised by the GP regression. Thus $\dot{\nu}$ values at greater distances from the minimum and maximum values affect the calculation of $\dot{\nu}$ in a particular window.  In tests where we constrained the covariance lengthscale to be 600 days, we found that $\Delta \dot{\nu} / \dot{\nu}$  was in better agreement with LHK. Lomb-Scargle periodicity analysis shows a strong peak at 0.290(2) yr\textsuperscript{-1}, correponding to a cycle time of 1260 days ($\sim$3.45 years).  

Our GP analysis reveals no evidence for any systematic variations in pulse shape over the timespan studied.  Additionally, we formed two total integrated profiles from all individual epochs where $\dot{\nu}$ was computed to to be strongest ($\dot{\nu} < -1.185 \times 10^{-15}$ Hz s\textsuperscript{-1}) and weakest ($\dot{\nu} > -1.175 \times 10^{-15}$ Hz s\textsuperscript{-1}) respectively, and formed their profile residual.  No significant differences were seen between the two. Therefore we find no evidence for pulse shape variations in this source.

%% file: sections/discussion/discussion.tex
\section{Discussion} \label{gprdisc}

We have used the GPR method developed by \cite{bkb+14} and BKJ to model the emission and rotational variability of the 17 pulsars originally studied by LHK. We have confirmed the $\dot{\nu}$ variability in all 17 sources and identified new transitions in 8 years of extended monitoring with wider bandwidth, improved cadence and higher time resolution. In all except 6 cases, we were able to model the timing residuals using a single squared exponential covariance function with an additional white noise covariance function. In the remaining cases, a second squared exponential covariance function was required to produce a satisfactory model. BKJ suggested that in these cases, there are two distinct underlying processes driving the modulations in the spin-down rate.

We have shown that the GPR method is capable of resolving extremely subtle pulse profile shape changes such as those in PSR B1540$-$06 (also seen in LHK) without having to specify a particular metric in advance such (e.g., specific pulse shape parameter). This was also demonstrated by BKJ using simulated pulse profile changes. However, in a number of pulsars in our sample (e.g., PSRs B1714$-$34, B1903$+$07, B1907$+$00, B1929$+$20) we were unable to identify any such variations. Given the somewhat lower quality of the available data for these sources (e.g., low cadence, low S/N), we would only expect to resolve pronounced pulse shape changes. Therefore we speculate that with longer integration times and an increase in the observing cadences, the Gaussian processes method may reveal equally subtle profile changes, such as those seen in PSR B1540$-$06, in those sources for which we currently observe no variations.  Additionally, the application of this method to observations recorded at other observing frequencies may also reveal emission variations, thereby allowing the broadband characterisation of mode-switching behaviour (see below), and flux calibrated profiles might reveal correlated flux variations with no pulse shape changes. 

We note that in many cases (e.g., PSRs B1540$-$06, B1642$-$03, B1828$-$11) that enhancement of smaller profile components occurs when $\dot{\nu}$ is in its lowest state (e.g., see Figure \ref{srccgroup}, panels C, D and F respectively). However, because our profiles are normalised by the mean on-pulse flux density (as flux calibrated data were not used in this study), enhancement of one component, necessarily results in reduction of another. This is particularly apparent in the variability maps for PSR B1828$-$11 (Figure \ref{1828_map}) where, when the precursor component is active, a decrement in main pulse occurs. This ``mirroring'' effect is an artifact of the method used and does not itself imply that the main-pulse flux rises. However, BKJ, using flux-calibrated data from the Parkes radio telescope, showed that in PSR B1828$-$11, the mean flux density when the pre-cursor is weak is 1.4 times greater than when the pre-cursor is strong. \cite{slk+18} also showed, using flux-calibrated data from the Green Bank Telescope, that the peak flux density is lower when the precursor is present. Other sources may exhibit similar behaviour and this could be identified by the use of flux-calibrated pulse profiles in a follow-up study. 

Our variability models have allowed us to confirm the mode-switching behaviour in 6 of the pulsars studied in LHK and that these mode-switches are contemporaneous with changes in the value of $\dot{\nu}$. In particular, as a result of 8 years of new data, PSR J2043$+$2740, which was observed to undergo just two large transitions in LHK, has undergone a further two similar transitions allowing a tentative estimate of its magnetospheric switching timescale of $\sim$10 years. Constrastingly, PSR B2035$+$36, which was seen to exhibit a single large increase in the magnitude of $\dot{\nu}$ (and a corresponding narrowing of the pulse profile) in 2004, has not undergone any further large transitions since. As a result we can only establish lower limits on the time B2035$+$36 spends in either of its stable states having spent 14 years to date in the high-$\dot{\nu}$ state and a minumum of 15 years in the low $\dot{\nu}$ state since monitoring began.  

We have observed correlated $\dot{\nu}$ and profile shape changes in PSR B1642$-$03 which is particularly conspicuous in the higher-time resolution DFB data (see Figure \ref{1642_map}). The magnitude of $\dot{\nu}$ gradually decreases (i.e., becomes less negative) over $\sim$2000 days towards a minimum value (see Figure \ref{nudots_prof1}). It then undergoes a much more rapid increase. The minima in $\dot{\nu}$ are clearly associated with a small increase in power in a weak leading edge component of the profile (Figure \ref{srccgroup}, panel D). \cite{lyne13} noted that a 20 per cent change in the ratio of the profile's cone and core components was seen to correlate with a $\sim$1 per cent change in $\dot{\nu}$. In addition, in the higher time-resolution DFB data, we have identified subtle variations in the trailing compononent of PSR B1826$-$17's triple-peaked profile that occur over timescales of 250-650 days without conforming to any clear periodicities. Additionally they do not appear to correlate with the value of $\dot{\nu}$ which is modulated on a timescale of $\sim$1100 days. This suggests that there may be magnetospheric changes in PSR B1826$-$17 that are too small to to affect $\dot{\nu}$ in a measureable way.

%% file: sections/discussion/discstates.tex
\subsection{Discrete magnetospheric states?}
\label{disc}

The variations in $\dot{\nu}$ can be interpreted as resulting from a global reconfiguration of the distribution of plasma in the pulsar magnetosphere. In intermittent pulsars such as PSR B1931$+$24, the open field line region is thought to become depleted of charged particles leading to cessation of radio emission and a reduction in the braking torque \citep{klo+06}. Whilst the pulsar is not emitting, the braking is mainly due to magnetic dipole radiation. Whilst emitting, additional braking results from the outflowing wind of particles that gives rise to radio emission. The pulsars studied here may be an analagous case, in which each emitting state of a pulsar is associated with a different density of plasma in the polar cap region that in turn gives rise to a distinct spin-down rate. 

\begin{table*}
\centering
    \caption[Magnetospheric and rotational quantities of the 17 pulsars used in a rotational/emission variability study]{\label{rotmag}Magnetospheric and rotational quantities of the 17 pulsars used in this work. $\dot{\nu}_{\mathrm{weak}}$ and $\dot{\nu}_{\mathrm{strong}}$ represent the most extreme values of $\dot{\nu}$ that each pulsar assumes. We show the change in magnetospheric charge density $\Delta \rho$ and the Goldreich-Julian density $\rho_{\mathrm{GJ}}$.  In the right-most column is the ratio of the two quantities expressed as a percentage. We also show the characteristic surface magnetic field strength $B_0$. $\Delta \rho$ values have been computed assuming a canonical moment of inertia of 10\textsuperscript{45} g cm\textsuperscript{2} and a neutron star radius of 10 km. Pulsars exhibiting correlated spin-down and profile variability are highlighted in gray.}
    \begin{tabular}{|lrrrrrrrr|}
    \hline
    PSR & \pbox{10cm}{$\nu$ (Hz)} & $\dot{\nu}_{\mathrm{weak}}$ (Hz s\textsuperscript{$-$1}) & \pbox{10cm}{$\dot{\nu}_{\mathrm{strong}}$ (Hz s\textsuperscript{$-$1})} & \pbox{10cm}{$\Delta \dot{\nu}$ (Hz s\textsuperscript{$-$1})} & $B_0$ (TG) & $\Delta \rho$ (mC m\textsuperscript{$-$3}) & \pbox{10cm}{$\rho_{\mathrm{GJ}}$ (mC m\textsuperscript{$-$3})} & $\Delta \rho / \rho_{\mathrm{GJ}}$ (\%) \\
     \hline
     \hline
\rowcolor{lightgray} B2035$+$36  &  1.62  &  $-$1.18e-14  &  $-$1.32e-14  &  1.4e-15 & 1.69  &  7.19  &  30.46  &  23.61 \\
\rowcolor{lightgray} B1822$-$09  &  1.30  &  $-$8.61e-14  &  $-$8.86e-14  &  2.5e-15 & 6.33  &  5.32  &  91.56  &  5.81 \\
\rowcolor{lightgray} J2043$+$2740  &  10.4  &  $-$1.32e-13  &  $-$1.40e-13  &  8.0e-15 & 0.35  &  4.81  &  40.50  &  11.89 \\
B1903$+$07  &  1.54  &  -1.14e-14  &  $-$1.22e-14  &  8.0e-16 & 1.79  &  4.29  &  30.67  &  14.00 \\
\rowcolor{lightgray} B1828$-$11  &  2.47  &  $-$3.64e-13  &  $-$3.66e-13  &  2.0e-15 & 4.97  &  1.50  &  136.59  &  1.10 \\
\rowcolor{lightgray} B0740$-$28  &  6.00  &  $-$6.02e-13  &  $-$6.06e-13  &  4.0e-15 & 1.68  &  1.51  &  112.16  &  1.34 \\
\rowcolor{lightgray} B1642$-$03  &  2.58  &  $-$1.17e-14  &  $-$1.19e-14  &  2.0e-16 & 0.83  &  0.82  &  23.83  &  3.46 \\
B1839$+$09  &  2.62  &  $-$7.40e-15  &  $-$7.60e-15  &  2.0e-16 & 0.65  &  1.02  &  18.95  &  5.39 \\
B1714$-$34  &  1.52  &  $-$2.27e-14  &  $-$2.29e-14  &  2.0e-16 & 2.57  &  0.77  &  43.46  &  1.77 \\
B1826$-$17  &  3.26  &  $-$5.85e-14  &  $-$5.89e-14  &  4.0e-16 & 1.31  &  0.65  &  47.52  &  1.38 \\
B1818$+$04  &  1.67  &  $-$1.77e-14  &  $-$1.78e-14  &  1.0e-16 & 1.97  &  0.41  &  36.61  &  1.13 \\
\rowcolor{lightgray} B0919$+$06  &  2.32  &  $-$7.37e-14  &  $-$7.4e-14  &  3.0e-16 & 2.45  &  0.52  &  63.24  &  0.82 \\
\rowcolor{lightgray} B1540$-$06  &  1.41  &  $-$1.73e-15  &  $-$1.76e-15  &  3.0e-17 & 0.79  &  0.44  &  12.39  &  3.51 \\
B1907$+$00  &  0.98  &  $-$5.31e-15  &  $-$5.35e-15  &  4.0e-17 & 2.4  &  0.40  &  26.17  &  1.51 \\
B2148$+$63  &  2.63  &  $-$1.17e-15  &  $-$1.19e-15  &  2.0e-17 & 0.26  &  0.25  &  7.61  &  3.33 \\
B1929$+$20  &  3.73  &  $-$5.86e-14  &  $-$5.88e-14  &  2.0e-16 & 1.08  &  0.30  &  44.82  &  0.68 \\
B0950$+$08  &  3.95  &  $-$3.57e-15  &  $-$3.60e-15  &  3.0e-17 & 0.24  &  0.18  &  10.55  &  1.73 \\
\hline
    \end{tabular}
\end{table*}

We use the method outlined in \cite{djw+18} (adapted for mode-switching pulsars from the method used in \cite{klo+06} to understand intermittent pulsar spin-down) to estimate the change in the magnetospheric charge density $\Delta \rho$ associated with the spin-down rate transitions,  with respect to the Goldreich-Julian density $\rho_{\mathrm{GJ}}$ \citep{gj69} for each pulsar. Table \ref{rotmag} shows the values of $\Delta \rho$, $\rho_{\mathrm{GJ}}$ and the ratio between them. Changes to the charge density $\Delta \rho$ are of the order of 0.1 to 10 mC m\textsuperscript{$-$3} corresponding to a fraction of $\rho_{\mathrm{GJ}}$ of the order of $\sim$1-25 per cent. We note a tendency for pulsars in which emission-rotation correlation is clearly observed, to have generally higher values of $\Delta \rho$. In particular, the largest value of $\Delta \rho$ is associated with PSR B2035$+$36 (approximately one quarter of $\rho_{\mathrm{GJ}}$) which correspondingly also shows the largest fractional change in $\dot{\nu}$. PSR B1903$+$07 shows charge density changes of approximately 14 per cent of $\rho_{\mathrm{GJ}}$, and may be an outlier in this respect, however it is perhaps no surprise that profile shape changes are not seen given the comparatively low quality of the available data on this source (low cadence, low S/N). Profile shape changes in PSR B1907$+$00 may be too modest to be detected with this method and data quality. Comparing the mean signal-to-noise ratios of the DFB profiles of PSR B1540$-$06 and PSR B1907$+$00, we find the former is $\sim$4 times greater than the latter.  Changes to the flux of the trailing edge component of PSR B1540$-$06's profile are of the order of 5 per cent of the peak flux density, therefore changes of similar magnitude to PSR B1907$+$00's profile may be resolved with a $\sim$4-fold improvement to the signal-to-noise. Though spin-down rate changes are associated with low $\Delta \rho$ values for PSRs B0919$+$06 and B1540$-$06, the average S/N of their profiles are notably greater than for PSRs B1839$+$09, B1714$-$34, B1826$-$17 and B1818$-$04, all of which have an apparently higher $\Delta \rho$ and yet have not been revealed to show profile shape changes associated with $\dot{\nu}$. 

Although a high value of $\Delta \rho$ may suggest that radio emission variations are occurring, it is not neccesarily the case that they should be detected at all frequencies. Though mode-switching is a broadband phenomenon, affected components are not necessarily visible at all frequencies.  For example, the pulse profile of PSR B1642$-$03 comprises a single component at frequencies lower than 150 MHz (\citealt{kl99}; \citealt{phs+16}) with the weak conal component that is seen to vary at L-Band (see \S \ref{1642}), being absent. Therefore a low-frequency search for pulse shape variations in this source is unlikely to reveal variability of the same nature identified in this work. However, the conal components become increasingly prominent at higher frequencies (\citealt{sgg+95}, \citealt{hx97}). PSRs B1818$-$04 and B1839+09 whose L-Band profiles comprise a single component, are amongst those that show no clear shape variations. However their profiles resolve into two distinct components at 4.85 GHz \citep{kkwj98}, whose shapes and/or relative intensities may vary in time, possibly coinciding with variations in $\dot{\nu}$, therefore similar analyses at other frequencies are beneficial. 

The variability of PSR B1822$-$09 is an interesting case study in the frequency dependence of mode-switching and the relationship between the radio emission and the conditions in the magnetosphere. The B-mode (in which the pulse profile has a greater flux density), corresponding to the appearance of the PC, is associated with the weaker spin-down state at L-band, which is the opposite of what would be expected by assuming the radio flux is proportional to the spin-down rate. At 325 MHz however \citep{bmr10}, this is reversed - with the B-mode corresponding to the stronger spin-down mode \citep{pj13}. A similar scenario is observed in the mode-switching of PSR B0826$-$34 \citep{elg+05}. \cite{hkh+17} characterised the pulsed thermal X-ray emission from PSR B1822$-$09 but did not identify any X-ray mode-switching behaviour, contemporaneous or otherwise, with the radio mode-switching, suggesting that the particles responsible for the X-ray emission do not play a role in regulating the spin-down of the star.  PSR B0943$+$10 is a similarly notable case in which contemporaneous radio and X-ray mode-switching \emph{is} observed, characterised by thermal X-ray pulsations exhibiting a greater flux in the radio Q-mode (\citealt{mkt+16}, \citealt{rmt+19}), suggesting that in this mode, the current flow is stronger. 


\cite{pj13} estimated the flux density changes of several mode-switching pulsars, concluding that we should expect to observe a similar number of cases of correlation and anti-correlation between flux density and $\dot{\nu}$ due to a combination of frequency dependence and line-of-sight effects. Considering the latter, they proposed that $\dot{\nu}$ is correlated with flux from the core region but anti-correlated with the conal region and so the nature of any observed correlation depends on where the line-of-sight crosses the pulsar's radio beam. For the 9 pulsars that exhibit no evidence of $\dot{\nu}$ correlated pulse shape variations, it may be the case that our line-of-sight does not cross an actively variable region of the radio beam.  There may also be intrinsic pulse intensity variations (i.e., the pulses retain the same shape but vary in flux) that correlate (or anti-correlate) with $\dot{\nu}$ which could be identified using flux calibrated pulse profiles. Moreover, the detection of profile shape and/or flux density changes that correlate with $\dot{\nu}$ may be symptomatic of, but not necessarily directly proportional to, the magnetospheric charge density which regulates a pulsar's spin-down. Low frequency observations using new generation telescopes, such as LOFAR and SKA-LOW\footnote{SKA-LOW refers to the low frequency component of the Square Kilometre Array, operating at a frequency range of 50-350 MHz.},  have a key role to play in understand the frequency dependence and the overall broadband nature of mode-switching. Observations at low frequencies will allow for a larger fraction of the pulsar emission beam to be sampled, which may reveal radio emission changes which are missed by monitoring at higher frequencies (e.g., \citealt{sha+10}, \citealt{kjk+15}).




\begin{figure}
    \includegraphics[width=1.0\columnwidth]{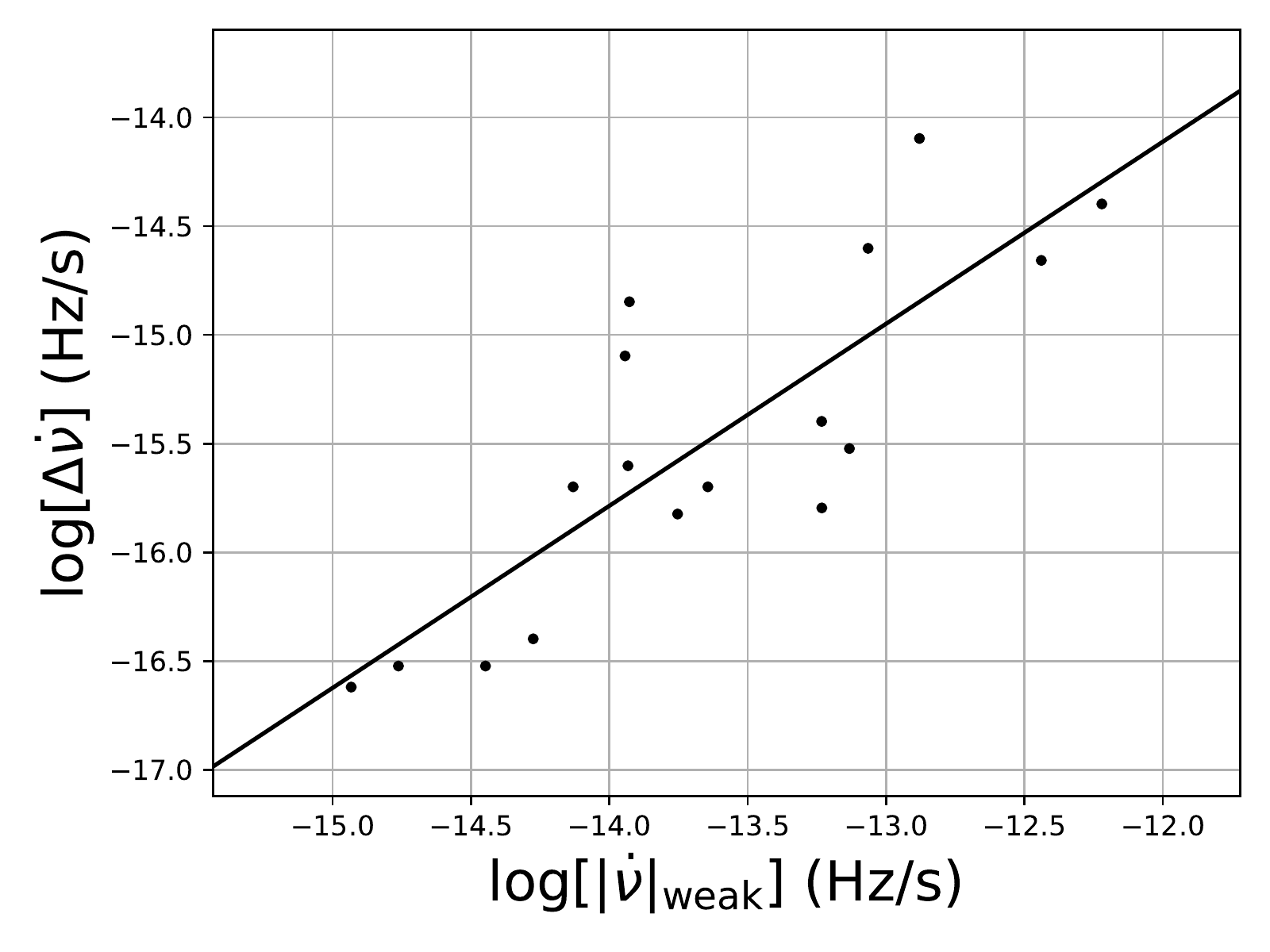} \\
    \includegraphics[width=1.0\columnwidth]{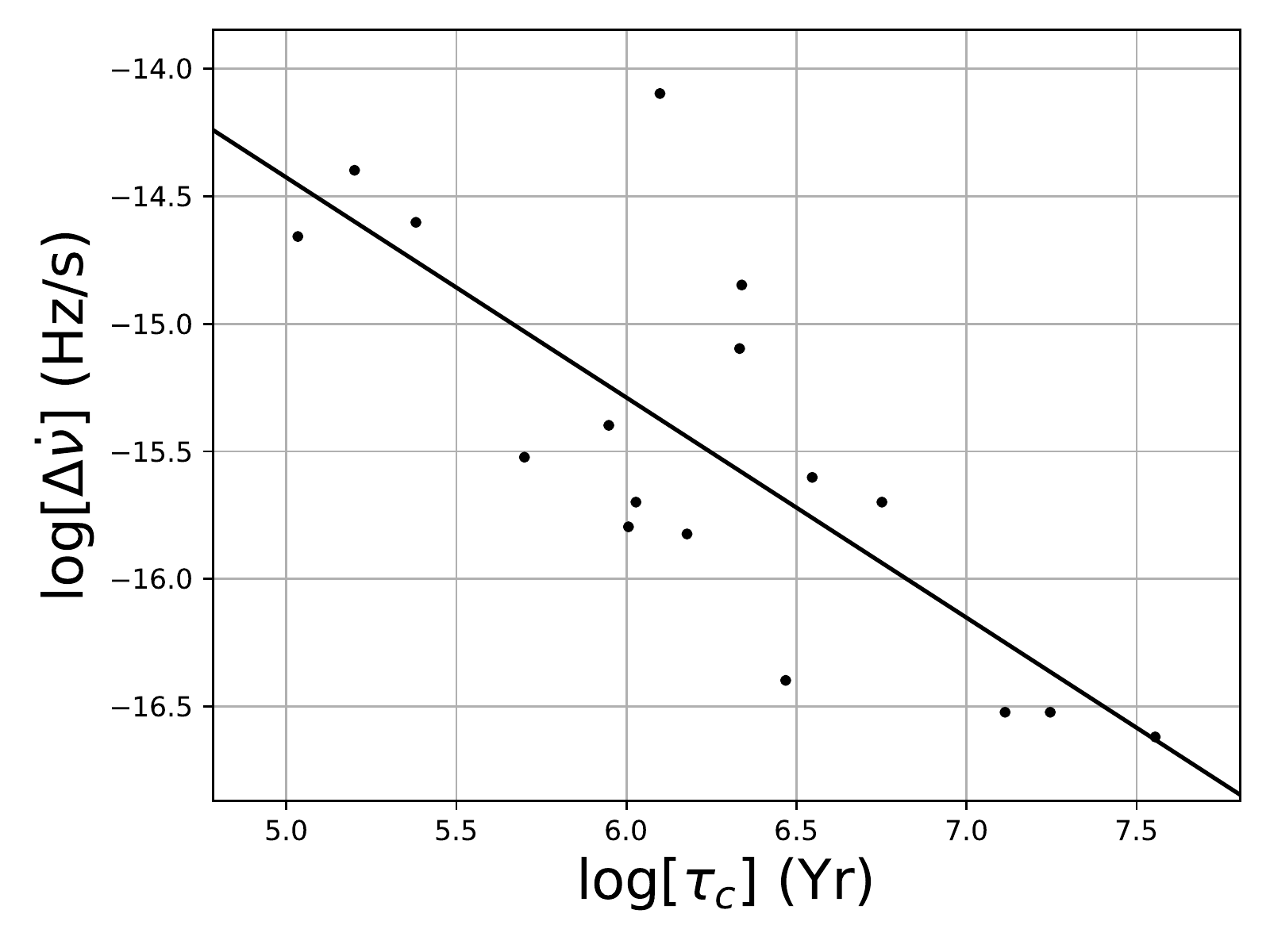}
    \caption[The dependence of the $\dot{\nu}$ transition amplitude on $\dot{\nu}$ and $\tau_c$]{\label{dnudot_corrs}The dependence of the change in the spin-down rate $\Delta \dot{\nu}$ of the 17 pulsars studied in this work, on the weakest value of $\dot{\nu}$ that they assume (upper panel) and the characteristic age $\tau_c$ (lower panel).  The Spearman Rank correlation coefficients $\rho_{\mathrm{srcc}}$ between the base 10 logarithms of the variables are +0.82 and -0.72 respectively for the upper and lower panels.  The black solid lines are straight line fits to the data and have slopes of $0.84 \pm 0.02$ (upper) and $0.86 \pm 0.04$.}  
\end{figure}

Figure \ref{dnudot_corrs} (upper panel) shows the dependence of $\Delta \dot{\nu}$ on the value of $\dot{\nu}$. A clear linear relation exists between the two quantities such that $\Delta \dot{\nu}$ is approximately 1 per cent of the spin-down rate. This was also noted in LHK (see LHK supplementary online material, Figure 10) using 68 pulsars.  Similarly, $\Delta \dot{\nu}$ is anticorrelated with the characteristic age (Figure \ref{dnudot_corrs}, lower panel), suggesting that $\dot{\nu}$ transitions are larger in younger pulsars (in which $\rho_{\mathrm{GJ}}$ is greater by virtue of having typically greater spin-frequencies).

%% file: sections/discussion/modtimes.tex
\subsection{Modulation timescales}

The modulation timescales in $\dot{\nu}$ vary strongly for different pulsar, from a few months (PSR B0740$-$28) to $\sim$10 years (PSR J2043$+$2740) or more (PSR B2035$+$36, PSR J0738$-$4042 \citep{krj+11}), indicating that magnetospheric state lifetimes assume a range of values across the population. We find no significant correlation between the modulation timescales (given by the position of the peaks in the Lomb-Scargle spectra; Figure \ref{LS}) and characteristic age, $\Delta \dot{\nu}$, $\nu$, $\dot{\nu}$, $B_0$, $\rho_{\mathrm{GJ}}$ or $\Delta \rho$.  We have noted that in many cases, the value of $\dot{\nu}$, can modulate on multiple timescales. For example, in addition to the single large transition observed in PSR B2035$+$36, significantly weaker and faster modulations occur in each of the otherwise stable states. With the possible exception of PSR J2043$+$2740, we have not observed any profile variations associated with these shorter timescale variations. However, these may be too modest to be seen with the available data quality. PSR B1818$-$04, which shows an approximately decade-long high-$\Delta \dot{\nu}$ modulation as well as shorter timescale (1.5 years) low-$\Delta \dot{\nu}$ modulations, exhibits no profile variations associated with either state.

PSRs B1828$-$11 and B1540$-$06 show exceptionally regular $\dot{\nu}$ modulations with peaks in the spin-down occurring every 500 and 1500 days respectively and this is reflected in the sharp dominant peaks in their Lomb-Scargle spectra. In many other cases however, the periodicities are notably less stable (e.g., PSRs B0740$-$28, B0919$+$06, B1642$-$03). The intermittent pulsar PSR B1931$+$24, which has been afforded high cadence monitoring since 2006, has been shown to cycle between distinct magnetospheric states on an average timescale of $\sim$38 days \citep{ysl+13}. It is less straightforward to constrain periodicities on timescales of a few days when cadences are irregular or highly infrequent as is the case for some pulsars in this sample. Identifying a large sample of state-switching pulsars has the potential to dramatically improve the statistics available on the nature of the switches and the relationships between the lengths of individual modes. Some progress has been made in this respect. \cite{khjs16}, using a large sample of 151 pulsars identified 7 cases where "nearly sinusoidal" modulations were present within the timing noise, leading to the suggestion that a highly periodic processes, such as precession, may regulate the magnetospheric switching.   

Though the modulations in $\dot{\nu}$ apparently occur on relatively long (months to many years) timescales, it has been shown, in some cases, that the switching rate of the pulse profile can be considerably more rapid (minutes, e.g., \citealt{slk+18}). As it is not possible to measure the value of $\dot{\nu}$ within such a short emission state (typically the values of $\nu$ at the start and end of an emission state are not sufficiently different to yield a measurement of $\dot{\nu}$, (e.g., \citealt{ssw18}), one is confined to measuring the average $\langle \dot{\nu} \rangle$ over some longer interval. Were $\dot{\nu}$ transitioning at a constant (though rapid) rate, we would not expect to measure $\langle \dot{\nu} \rangle$ modulations over any timescale as the average value would be identical regardless of where in time one chooses to measure it. However, the fact that we observe longer timescale spin-down rate modulations in these sources indicates that the average value is changing over time and that the value at any given time is a consequence of the fraction of time, over some averaging timescale, that the pulsar spends in one state or the other. The $\dot{\nu}$ modulations observed therefore, are the result of a slowly changing mixture of the two states (see LHK). The fact that the evolution of $\dot{\nu}$ closely traces the pulse shape suggests that $\dot{\nu}$ is switching on an equally short timescale but individual transitions are too frequent to be resolved in these sources.

%% file: sections/discussion/other.tex
\subsection{Other models}

\textcolor{black}{The modulations in PSR B1828$-$11 have been interpreted as the pulsar freely precessing. In this model the spin and angular momentum vector are not parallel resulting in the impact parameter ($\beta$) changing with time, leading to gradual, highly periodic changes to the pulse shape as well as a variable torque on the pulsar. Clearly free precession alone cannot account for the entire phenomenon of correlated emission and spin-down variations as generally, the modulations do not have sufficiently well-defined periodicities. Comparing switching and precession models for PSR B1828$-$11 using a Bayesian framework, \cite{ajp16} concluded that the precession model was somewhat favoured over the switching model when considering the long-term ($\sim$500 days) observed modulations. The precession model of PSR B1828$-$11  is difficult to reconcile when considering the short term modulations that have been shown to ultimately result in the "apparent" longer term modulations (see \cite{slk+18} for a recent review of the modulations in this pulsar).  It has also been argued that the precession and switching models are not necessarily mutually exclusive, with precession regulating the switching timescale (\citealt{jones12}, \citealt{khjs16}, \citealt{ajp17}). However, the highly periodic modulations in PSR B1828$-$11 were not affected by the occurrence of its 2009 glitch as expected by precession models (e.g., \citealt{jap17}).}

\textcolor{black}{The single large transition observed in PSR B2035$+$36 has interesting parallels with PSR J0738$-$4042 (B0736$-$40) in which a single, isolated transition in both profile shape and $\dot{\nu}$ occurred in 2005. \cite{bkb+14} interpret this as evidence that the pulsar's ordinary magnetospheric state was interrupted by an encounter with an asteroid. In this scenario, an asteroid entering the magnetosphere, is evaporated and ionised and the injection of charges results in disruption of the existing cascade processes above the polar cap, leading to a change in the pulse shape with a corresponding change to the rate of particle outflow and, consequently, the spin-down rate. Eventually, it may be the case that the pulsar returns to its previous state when the injected fuel is exhausted. In this sense, if the PSR B2035$+$36 event was due to an encounter with an asteroid, then it may, in time, return to its prior \emph{preferred} state. Future data may reveal this to be the case or in fact reveal a very long timescale periodicity in PSR B2035$+$36's transitioning behaviour, due to magnetospheric switching or repeated encounters with circumstellar material. The strong periodicities observed in many other pulsars (e.g., PSRs B1828$-$11 and B1642$-$03) are similarly incompatible with this model as periodic encounters with similar mass asteroids from a fossil disk would be required to explain the regularity of the modulations.} \cite{kyw+18} suggested that the transition in PSR B2035$+$36 is due to a small glitch with an unusually large and persistent change in $\dot{\nu}$, proposing that magnetospheric changes may be driven by conditions in the neutron star interior.
  
PSR B1822$-$09 is known to mode-switch on timescales of several minutes between two modes defined by the relative and absolute changes in intensity between a precursor component and an interpulse component. This pulsar has exhibited a unique phenomenon within our sample in that between MJDs 50000 and 55000 it underwent several sporadic events where $\dot{\nu}$ became temporarily weaker for short (up to 200 days) periods of time before reverting back to stronger spin-down.  We have confirmed that the power in the precursor component became enhanced during these episodes suggesting that the pulsar was spending an increased fraction of its time in the B-mode before reverting back to a more uniform transition timescale. We note that since MJD 55000 the pulsar has not undergone any further such events. 

A number of our pulsars demonstrate $\dot{\nu}$ evolution curves that comprise consecutive major and minor peaks (Figure \ref{nudots_prof1}). This is particularly apparent in PSR B1828$-$11 and to a lesser extent in PSRs B0919$+$06, B1642$-$03 and B1907$+$00.  Whilst multi-peaked modulations could indicate that the pulsar is able to assume more than two states with different $\dot{\nu}$ values, \cite{psw14} showed that such modulations can arise from just two $\dot{\nu}$ values if the distribution of time spent in each state is bimodal. They were able to reproduce the double-peaked $\dot{\nu}$ modulations seen in PSR B0919$+$06 using a sliding boxcar over many repeats of the intrinsic $\dot{\nu}$ variability in the figure. PSR B1642$-$03 shows mutiple $\dot{\nu}$ peaks that gradually increase in amplitude corresponding to larger and larger changes until they reach a maximum. Following this the pattern repeats. This may be a manifestation of the same behaviours seen in PSRs B1828$-$11 and B0919$+$06, in which the pulsar gradually spends larger periods of time in the low $\dot{\nu}$ state, though we offer no explanations as to why a pulsar magnetosphere would behave in this way. 

To date, only a handful of pulsars have been shown to exhibit correlated emission and spin-down variabililty.  We speculate that all pulsars which show timing noise may also be experiencing similar variability though this has so far remained unobserved. This is because the data from many of these pulsars have not been systematically explored for such variations or that the variations are too slight to be seen given the available data quality, the variations are occuring in regions of the beam not observable from Earth or that insufficient frequency coverage has been accrued.  Of the large samples of pulsars studied by LHK, BKJ and \cite{khjs16}, they respectively revealed 6, 9 and 1 source(s), to exhibit coincident profile/rotational variations (though in the latter study, two further pulsars showed promising hints of correlated profile changes).  A large number of pulsars are known to exhibit timing noise. Future work should attempt to identify/rule out profile and rotational variations in all of these sources as in many cases, long time baselines have already been accrued.  In addition, new generations of telescopes (e.g., the SKA, MeerKAT, CHIME, UTMOST) are set to dramatically increase the number of pulsars known to exhibit state-switching thereby allowing a fuller understanding of state-switching and its prevelance within the pulsar population. Furthermore, the ability to predict the spin-down rate from observations of the emission state may allow the effects of timing noise to be mitigated, thereby improving the utility of pulsars as precision timing tools.

%% file: sections/conclusions/conclusions.tex
\section{Conclusions}
\label{conc}

We have expanded upon the work of \cite{lhk+10} by analysing the spin-down and pulse shape variations in 17 pulsars whose timing residuals exhibit timing noise. Using a Gaussian processes technique developed in \cite{bkb+14} and \cite{bkj+15} we have confirmed the emission and rotational variations therein and revealed new transitions in 8 years of extended monitoring.  We conclude the following

\begin{itemize}
   \item{We have confirmed the correlated emission and spin-down variability previously noted in PSRs B0740$-$28, B1540$-$06, B1822$-$09, B1828$-$11, B2035$+$26, J2043$+$2740 (LHK) as well as PSR B0919$+$06 (\citealt{psw14}).}
   \item{We have characterised the correlated emission and spin-down variability in PSR B1642$-$03 in which a subtle precursor becomes marginally brighter when the magnitude of the spin-down is small.}
   \item{In 9 pulsars from our sample, though significant $\dot{\nu}$ variations occur, no correlated pulse shape changes were identified.  We consider the possibility that either a) they are not occurring (at least at the specific observing frequency used here), b) they are occurring but are too modest to be observed or c) they are occurring in a region of the beam that does not cross the line-of-sight towards the Earth. Longer, higher cadence, and multi-frequency observations could address possibilities (a) and (b). It is also possible that instead of pulse shape changes, flux density changes are occurring that correlate with $\dot{\nu}$. This could be investigated by the use of flux calibrated pulse profiles.}
   \item{We have shown that Gaussian process regression is able to resolve extremely subtle profile shape variations if the data quality is sufficiently high. We have noted that pulsars that show clear pulse shape changes that are correlated with $\dot{\nu}$ have a tendency to undergo larger $\Delta \rho$.  This potentially suggests that pulsars which undergo large charge density changes but have not been shown to exhibit profile variability should be afforded longer observations and multiple frequencies in order to reveal any shape changes.}
\end{itemize}